\documentstyle[emulateapj,onecolfloat,psfig]{article}

\newcommand{\da}{d_A}

\newcommand{\veck}{{\bf k}}
\newcommand{\vecl}{{\bf l}}

\newcommand{\tableskip}{\\[-6pt]}
\newlength{\tskip}
\newlength{\colwidth}

\newcommand{\beq}{\begin{equation}}
\newcommand{\eeq}{\end{equation}}
\newcommand{\beqa}{\begin{eqnarray}}
\newcommand{\eeqa}{\end{eqnarray}}

\newcommand{\bn}{\hat{\bf n}}

\newcommand{\rad}{r}    

\begin{document}
\twocolumn[
\title{First Star Signature in Infrared Background Anisotropies}
\author{Asantha Cooray$^{1}$, James J. Bock$^{1,2}$, Brian Keating$^{1}$, Andrew E. Lange$^{1}$, T. Matsumoto$^3$}
\affil{$^1$Division of Physics, Mathematics and Astronomy, 
California Institute of Technology, MS 130-33, Pasadena, CA 91125\\
$^2$Jet Propulsion Laboratory, MS 169-327, 4800 Oak Grove Drive, Pasadena, CA 91109\\
$^3$Institute of Space and Astronautical Science, Yoshinodai 3-1-1, Sagamihara,Kanagawa 229 8510, Japan}

\begin{abstract}

Recent cosmic microwave background anisotropy results from the Wilkinson Microwave Anisotropy Probe suggest that the universe was reionized
at a redshift around 20 with an optical depth for Thomson-scattering of 0.17 $\pm$ 0.04.  Such an early reionization 
could arise through the ionizing radiation emitted by metal-free population III stars at redshifts of 10 and higher.
We discuss infrared background (IRB) surface brightness spatial fluctuations
from such a generation of early star formation. We show that the spatial clustering of these stars
at tens of arcminute scales generates a contribution to the angular power spectrum of the IRB anisotropies
at the same angular scales. This excess can be potentially detected when
resolved  foreground galaxies out to a redshift of a few is removed from the clustering analysis.
We do not expect faint galaxies at redshifts of $\sim$ 3, with magnitudes less than 20 in the K-band, 
to be a source of strong confusion, since the fractional contribution  to the IR
background from these galaxies is at a level less than a few percent, while the expected contribution from first stars
 can be more than 50\% or more.
Additionally, assuming a population III stellar spectrum, we suggest that the clustering excess related to the first generation of stars
 can be separated from brightness fluctuations resulting from other foreground sources and galaxies
using multifrequency observations in the  wavelength range of $\sim$ 1 to 5 $\mu$m. In addition to removing the low-redshift
galaxy population, the multifrequency data are essential to account for 
certain foreground contaminants such zodiacal light, which, if varying spatially over degree scales, can be a significant source of
confusion for the proposed study.
Using various instruments, we study the extent to which spatial fluctuations of the IRB can be studied in the near future.
\end{abstract}

]

\section{Introduction}

The Wilkinson Microwave Anisotropy Probe (WMAP) has now provided strong evidence for an optical depth for electron scattering
of 0.17 $\pm$ 0.04 based on the large scale polarization pattern related to rescattering of Cosmic Microwave Background (CMB) 
photons (Kogut et al. 2003). If the reionization process is described as instantaneous and homogeneous,
the measured optical depth implies a reionization redshift of $\sim 17 \pm 5$ in a spatially flat universe.
Such a reionization redshift is higher than previously suggested by observational data involving the 
presence of a Gunn-Peterson trough  in $z \sim 6$ quasars (Fan et al. 2002). The derived redshift for reionization 
is at the high end of expectations related to reionization scenarios based
on the ionizing radiation from the first generation of 
star formation associated with population III (hereafter Pop III) stars (e.g., Cen 2003; Fukugita \& Kawasaki 2003; 
Venkatesan, Tumlinson \& Shull 2003; Wyithe \& Loeb 2003). 
These stars are expected to be very massive 
with a top-heavy mass function (Bromm, Coppi \& Larson 1999, 2002; Abel, Bryan \& Norman 2000, 2002), and their detection
is important to understand the astrophysics associated with  the reionization process that followed the end of dark ages.

While the direct detection of an individual Pop III star or a star cluster is beyond the capability of upcoming
telescopes or instruments, the first star signature can be potentially detected via indirect methods such as
associated line emission, e.g., HeII recombination lines of the surrounding ionized halo 
(Oh, Haiman \& Rees 2001; Tumlinson, Giroux \& Shull 2001) or through Ly-$\alpha$ emission lines (e.g., Tumlinson, Shull
\& Venkatesan 2003).
In Oh, Cooray \& Kamionkowski (2003), it was suggested that the heating of the surrounding interstellar medium by 
Pop III supernovae, and the subsequent transfer of energy to the CMB through Compton cooling,
 lead to a  substantial contribution to small angular scale CMB anisotropies, similar to the Sunyaev-Zel'dovich (SZ; Sunyaev \& Zel'dovich 1980) effect.
While an individual Pop III supernova remains  unresolved and undetectable, 
the large bias factors of halos containing these supernovae, with respect to the linear
density field, lead to an excess clustering signature in CMB anisotropies at tens of  arcminute scales 
when compared to the shot-noise contribution related to their finite density. 
The CMB anisotropies related to Pop III supernovae 
can be separated from the dominant SZ
fluctuations  related to galaxy  clusters since massive clusters that dominate the angular power spectrum at arcminute scale
can be identified and removed in future higher angular resolution CMB data (Cooray et al. 2003). 

Here, we suggest that a signature of stars themselves can be found in the infrared background (IRB).
In particular, if Pop III stars are found primarily at redshifts between 10 and 30, they are expected to contribute 
to the IRB at wavelengths between 1 and 5 $\mu$m.  Recent estimates now suggest that a large fraction of the IRB total intensity may be due to
these stars (Santos, Bromm \& Kamionkowski 2002; Salvaterra \& Ferrara 2003);
A substantial IRB arises from the Pop III population not only due to the direct emission associated with these stars, but also due to 
indirect processes that lead to free-free and Lyman-alpha emission from the ionized nebulae, or HII regions, surrounding  these stars.
The fractional contribution to the absolute, and isotropic, background by the Pop III population, however, cannot 
be easily determined observationally due to additional contributions
from foreground galaxies, stars, and  confusion from zodiacal light. 

Instead of the background intensity, or the monopole, 
we suggest that the presence of first stars can be established through spatial fluctuations in the surface brightness of the
IRB. For this purpose, we make use of the angular power spectrum of IRB anisotropies and show that the
power spectrum is expected to contain an excess clustering signature at tens of arcminute scales related to the Pop III population. 
While Pop III stars may dominate tens of arcminute-scale fluctuations
at short IR wavelengths (between 1 and 3 $\mu$m), any unresolved galaxies at redshifts of 3 and higher,
are expected to dominate IRB anisotropy  at wavelengths greater than 4 $\mu$m 
over all angular scales of interest. Thus, we expect the Pop III 
clustering signature can be separated from the angular power spectrum due to foreground
galaxies and other sources in multifrequency near-IR (NIR) images due to its unique spatial and spectral signature.
Furthermore, when the angular power spectrum due to Pop III stars is estimated, one can use that information, for example,
to measure the formation rate of the first generation of stars in the universe 
at redshifts between 10 and 30.

Note that spatial fluctuations in the IRB have already been detected in several ground-, suborbital- and space-based experiments:
Xu et al. (2002) at 4 $\mu$m using a rocket-borne experiment; at J, H and K-band with 2MASS data by Kashlinsky et al. (2002; also Odenwald et al. 2003); 
Infra-Red Telescope in Space (IRTS) data by Matsumoto 2000; and at large scales with Diffuse Infrared Background Experiment (DIRBE) on 
Cosmic Background Explorer  (COBE) by Kashlinsky, Mather \& Odenwald (1996; also, Kaslinsky et al. 1996). 
Absolute background measurements, generally indicate  an 
excess above that expected from galaxies alone (Kashlinsky \& Odenwald 2000; Madau \& Pozzetti 2000; 
Cambr\'esy et al. 2001; Wright \& Johnson 2001).  
Additionally, IRTS data also indicate a clustering excess at 100 arcmin scales between 1.4 to 2.1 $\mu$m, 
which remains unexplained by the known properties of low redshift galaxies (Matsumoto 2000,2001).
Expectations for an overall excess in anisotropy include both faint galaxies at high redshifts (Jimenez \& Kashlinsky 1999)
as well as a Pop III component related to the first generation of stars at redshifts greater than 10 
(e.g., Magliocchetti, Salvaterra \& Ferrara 2003). 
While simple estimates suggest that the IRB total intensity can be easily explained with Pop III stars alone (Santos, Bromm \& 
Kamionkowski 2003; Salvaterra \& Ferrara 2003), it is useful to also consider if these stars can be used to explain  anisotropy fluctuations
already detected or, if not, the expected level of spatial fluctuations from these stars.

The current  observational results related to IRB fluctuations are, unfortunately, 
limited to either small angular scales  (such as $\sim$ 1 to 30 arcseconds
measurements by 2MASS) or large angular scales ($>$ 0.5 degrees with COBE DIRBE), with limited 
measurements at arcminute to tens of arcminute scales.  While 2MASS measurements by Kashlinsky et al. (2003) are useful to
understand the presence of a Pop III contribution, at such small angular scales, both
foreground sources and Pop III stars  produce power-law like contributions to the angular power spectrum
and cannot easily be separated (Magliocchetti, Salvaterra \& Ferrara 2003). 
Additionally, the Pop III contribution can also be misidentified easily as a source of noise,
since, at these small angular scales, one expects a shot-noise type power spectrum due to the finite number density of these stars.

In order to understand the clustering nature of the Pop III population, here
we study the angular power spectrum of IRB fluctuations, as a function of wavelength, and introduce the use of
power spectra at each of these wavelengths and the cross-spectra between different wavelengths as a basis to 
study the star formation history related to Pop III objects. Our calculations extend those of Magliocchetti, Salvaterra \& Ferrara (2003), by considering the angular power spectrum of Pop III stars over a wide range of angular scales instead of  arcsecond
scales studied there with respect to 2MASS measurements. We argue that a well-planned observational program, concentrating on the
clustering at arcminute to tens of arcminute scales or more in the wavelength range of 1 to 5 $\mu$m, is
clearly needed to understand the presence of Pop III stars in IRB data. Since Pop III stars are expected at redshifts below
30, we find that the upcoming and planned
wide-field IR  and near-IR missions from space, such as SIRTF\footnote{http://sirtf.caltech.edu/} and the proposed WISE mission,
are unlikely to play a major role since the Pop III signature is reduced for $\lambda \geq 3$ $\mu$m. 
The upcoming ASTRO-F\footnote{http://www.ir.isas.ac.jp/ASTRO-F}
mission has imaging capabilities at arcsecond resolution in the K, L and M-bands between $\sim$ 1.7 to 5 microns with its Near Infra-Red Camera (N-IRC)
and at a higher sensitivity at short wavelengths (Watarai et al. 2000; Pearson et al. 2001).
While observations  with ASTRO-F are only restricted to the 10$' \times 10'$ field of view, 
these observations are still useful to understand the presence of Pop III sources  and their contribution to IRB spatial fluctuations 
at angular scales of few arcminutes. 

In order to understand the spatial fluctuations due to Pop III sources at tens of arcminute scales and above, and also to
probe clustering at arcminute scales at wavelengths below 2 microns, we also consider a multiwavelength wide-field survey using a rocket-borne
experiment. This experiment is planned to image the sky in multiple bands in the 
wavelength range between 1 and 5 $\mu$m, with the primary goal of exploring the presence of Pop III stars in
IRB spatial fluctuations. While we present a general discussion, we will focus on the ability of these experiments to extract
Pop III information through spatial clustering and to remove
confusing point sources related to both galactic stars and foreground galaxies.

The paper is organized as follows. In the next section, we calculate the power spectrum of IRB anisotropies both for
Pop III stars and foreground sources. To describe the emission from Pop III stars, we make use of the stellar spectra
calculated by Santos, Bromm \& Kamionkowski (2003). These calculations also include nebular emission associated with both free-free and Lyman-$\alpha$ radiation associated from the ionized HII region surrounding individual Pop III stars. 
Note that though we consider Pop III sources as the source or ionization, our calculations does not necessarily depend on this assumption. 
The WMAP results indicate evidence for a highly efficient source of starformation at high redshift and could be
in the form of Pop II to Pop III stars. While we make use of a Pop III spectrum to illustrate the clustering in spatial
fluctuations in the IRB, we also expect similar spatial fluctuations (though with a different amplitude) if the reionization is
related to a different stellar spectrum, say associated with mildly metal-enriched stars, instead of pure metal-free stars.
In such a scenario, one can simply replace the correct stellar spectrum instead of the Pop III one used here and repeat our calculations.
When illustrating our calculations, we take cosmological parameters from the currently favored $\Lambda$CDM 
cosmology consistent with recent WMAP results (Spergel et al. 2003), including a normalization for the matter power spectrum, at scales of 8 $h^{-1}$ Mpc,
of $\sigma_8=0.9$.

\begin{figure}[t]
\psfig{file=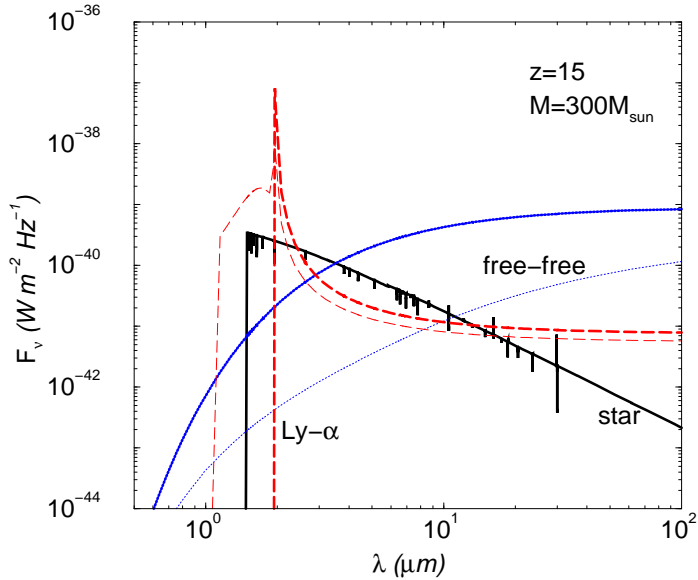,width=3.6in,angle=-90}
\caption{The emission spectrum of a 300 $M_{\rm sun}$ Pop III star at a redshift of 15. In addition to the
stellar spectrum (solid line), we also show the nebular 
emission related to Lyman-$\alpha$ (dashed lines) and free-free (dotted lines) of the ionized HII region surrounding the star. This nebular emission
depends strongly on the escape fraction of ionizing photons from the nebula, $f_{\rm esc}$,
 and we consider two extreme cases with $f_{\rm esc}=0$ (top lines) and 1 (bottom lines), respectively. 
In the redshift range of 10 to 30, the spectrum peaks at wavelengths of order 1 to 3 $\mu$m, where the study of
Pop III stars is strongly favored.}
\label{fig:spec}
\end{figure}

\section{Angular Power Spectrum of IRB fluctuations}
\label{sec:fisher}

\begin{figure*}[t]
\centerline{\psfig{file=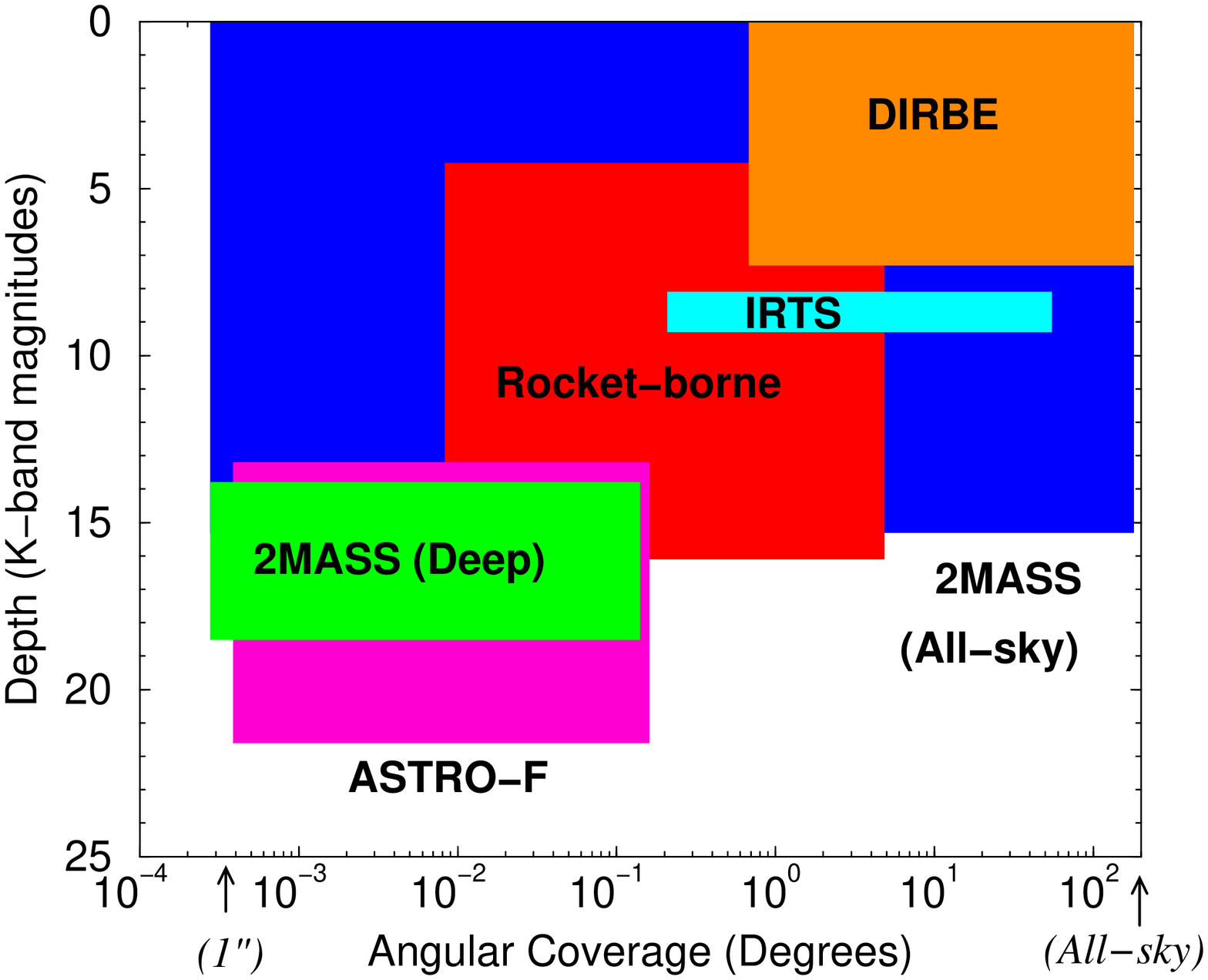,width=3.0in,angle=-90}\psfig{file=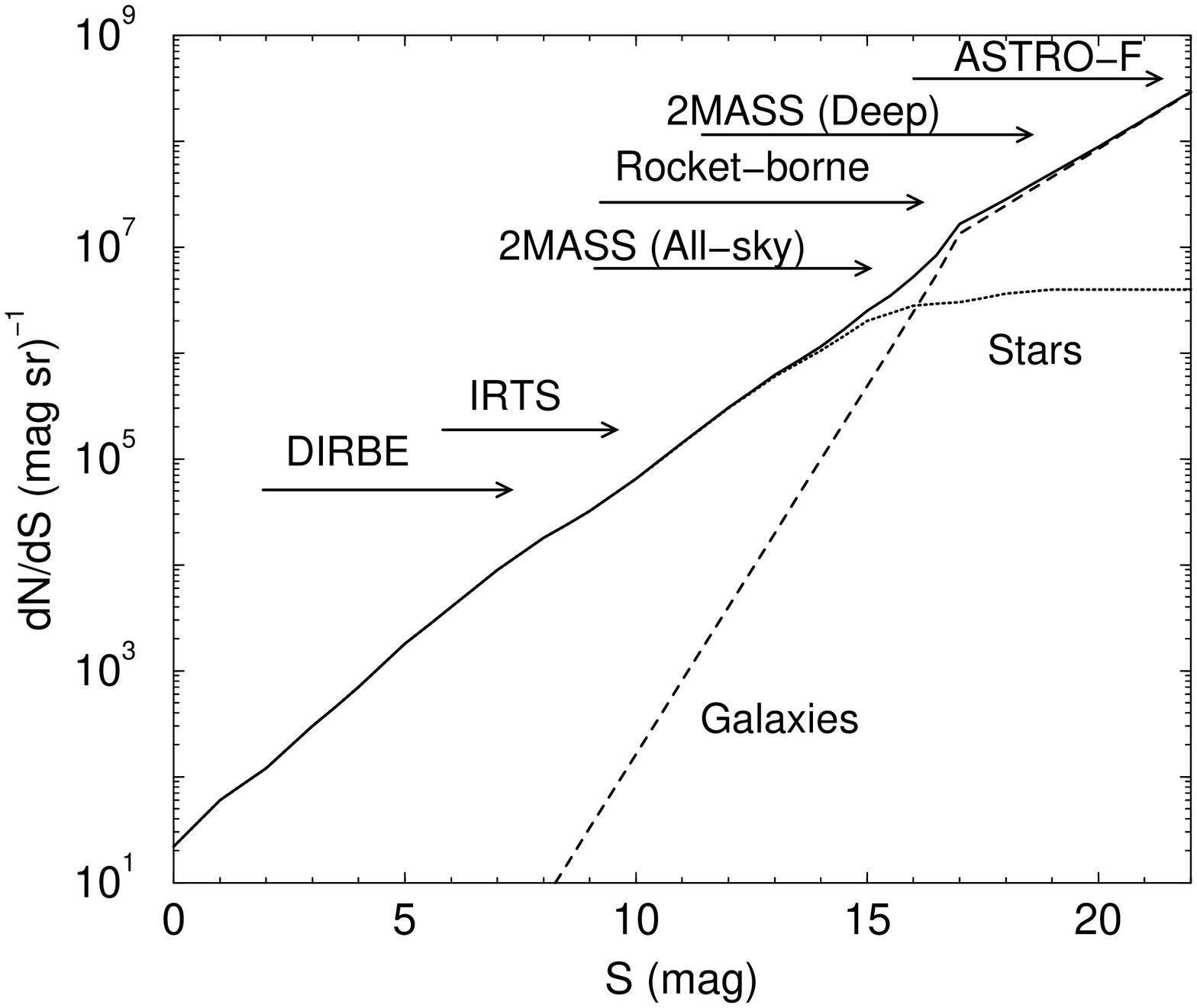,width=3.0in,angle=-90}}
\caption{{\it Left:} A summary of past or upcoming K-band imaging surveys related to clustering studies. The plotted bands represent the
one-sided sky coverage (with the minimum set at the resolution scale and maximum set at the field of view) in degrees (x-axis) and 
the survey depth between the K-band magnitude at which 10 sources per sky area covered are found down to the
5 $\sigma$ detection limit for resolved sources (y-axis). To detect Pop III clustering, surveys that populate the lower left hand corner
(large sky-area and large depth) are generally preferred, though a combination of experiments, 
such as a wide-field rocket-borne instrument and the ASTRO-F, may provide a first opportunity. {\it Right:}
Average K-band counts from a variety of sources for both galaxies and stars (based on Saracco et al. 2001). 
The arrows indicate the 5 $\sigma$ magnitude limit to which past or upcoming clustering measurements at K-band can be used to remove foreground sources
which are resolved in the data, including both instrumental noise and confusion (c.f., Lagache, Dole \& Puget 2003 for a description on how to
calculate the noise based on source counts). 
In the case of DIRBE, while sources are only resolved down to a higher flux, the clustering
analysis accounts for most stars based on detailed modeling of the Galactic star distributions (see, Kashlinsky et al. 1996 for details).}
\label{fig:counts}
\end{figure*}

Following standard approaches in the literature, we calculate the angular power spectrum of IRB surface brightness fluctuations
both due to foreground sources at redshifts between 0 and 7 and Pop III stars at redshifts between 10 and 30;
except for the lower end of the Pop III distribution (which we simply take here as the transition from
Pop III to Pop II stars, and discussed further in Section~2.1), our calculations are independent of the end points of the redshift ranges considered.

In general, the clustering contribution to the angular power spectrum resulting from foreground galaxies is mostly a 
power-law, which is now well explained with techniques such as the halo model (e.g., Cooray \& Sheth 2002).
Pop III stars, or any other source that trace linear clustering at redshifts of order 15 or so, on the other hand, 
have an angular power spectrum that peaks at tens of arcminute scales. 
Such a spectrum is expected since high redshift sources are expected to be strongly biased with respect to the linear dark matter density field
and linear fluctuation power spectrum contains a peak, or a turnover, at the scale corresonding to the matter-radiation equality.
A detection of such a peak in the power spectrum of any tracer source of linear density field would be significant since it will
help establish our basic uncerstanding of clustering evolution. Moving to smaller scales corresponding to  few tens of
 arcseconds and below, one expects the background Pop III stars to show a 
shot-noise type spectrum associated with the finite number density of these sources on the sky.
In the case of Pop III stars,  the transition scale  from linear to shot-noise type clustering is highly model dependent,
though with observational data, one can use a measured transition as an additional constraint on the
astrophysics of the Pop III population. Given that shot-noise contributions result from
both detector noise and other confusions, such as galactic foregrounds, we do not expect a precise measurement
of the shot-noise power associated with Pop III stars to be feasible. For similar reasons, it is unlikely that
arcsecond-scale IRB clustering information is useful for a separation  of the Pop III contribution from
foreground galaxies.

Our calculational approach  to model the angular power spectrum of the IRB 
is similar to the one that was 
used in Knox et al. (2001) to understand the clustering in the far-infrared background at wavelengths 
of few hundred microns or more due to dusty star burst galaxies at redshifts of $\sim$ 3.
The contribution to the IRB intensity, at a given wavelength and towards a 
direction $\bn$,  can be written as a product of the mean IRB emissivity and its fluctuation
\begin{equation}
I_\lambda(\bn) = \int_0^{z_{\rm max}} dz\; \frac{dr}{dz} a(z) \bar{j}_\lambda(z) \left[1+\frac{\delta j_\lambda(r(z)\bn,z)}{\bar{j}_\lambda(z)}\right] \, ,
\end{equation}
where $\bar{j}_\lambda(z)$ is the mean emissivity per comoving
unit volume at wavelength $\lambda$ as a function of redshift $z$ and
$\rad$ is the conformal distance or lookback time, from the observer, given by
\begin{equation}
\rad(z) = \int_0^z {dz' \over H(z')} \,,
\end{equation}
where the expansion rate for adiabatic cold dark matter
cosmological models with a cosmological constant is
\begin{equation}
H^2 = H_0^2 \left[ \Omega_m(1+z)^3 + \Omega_K (1+z)^2
+\Omega_\Lambda \right]\,.
\end{equation}
Here, $H_0$ can be written as the inverse
Hubble distance today $cH_0^{-1} = 2997.9h^{-1} $Mpc.
We follow the conventions that in units of the critical density $3H_0^2/8\pi G$,
the contribution of each component is denoted $\Omega_i$, $i=c$ for the CDM, $b$ for the baryons,
$\Lambda$ for the cosmological constant.
We also define the auxiliary quantities $\Omega_m=\Omega_c+\Omega_b$ and $\Omega_K=1-\sum_i \Omega_i$,
which represent the matter density and the contribution of spatial curvature to the expansion rate respectively.

The absolute IRB has now been studied both
observationally  and theoretically (e.g., Hauser \&  Dwek 2001 for a recent review). Here, we focus on the spatial fluctuations of
the background, $\delta I_\lambda(\bn)$, and consider the angular power spectrum of the IRB, which is simply the Legendre
transform of the two point correlation function, $C(\theta)$:
\begin{equation}
C_l = 2\pi \int \theta \; d\theta C(\theta) J_0(l \theta) \, .
\end{equation}
Denoting the Fourier transform of $\delta I_\lambda(\bn)$ as $\delta I_\lambda(\vecl)$, one can define the angular power spectrum 
of the IRB at wavelengths $\lambda$ and $\lambda'$, in the flat-sky approximation, as 
\begin{equation}
\langle \delta I_\lambda(\vecl) \; \delta I_\lambda(\vecl') \rangle= (2\pi)^2 \delta_D(\vecl+\vecl') C_l^{\lambda \lambda'} \, .
\end{equation}
In order to calculate the spatial fluctuations related to the emissivity, 
we assume $\delta j_\lambda(r(z)\bn,z)/\bar{j}_\lambda(z)$ trace fluctuations in the source density field,
$\delta_s = \delta \rho_s/{\bar \rho_s}$, such that, in Fourier space,
\begin{equation}
\frac{\delta j_\lambda(\veck,z)}{\bar{j}_\lambda(z)}= \delta_s(\veck,z) \, .
\end{equation}
The source density field fluctuations are defined by the three dimensional power spectrum, which we define as
\begin{equation}
\langle \delta_s(\veck,z) \delta_s(\veck',z) \rangle = (2\pi)^3 \delta_D(\veck+\veck')P_{ss}(k,z) \, .
\end{equation}

\begin{figure*}[!ht]
\centerline{\psfig{file=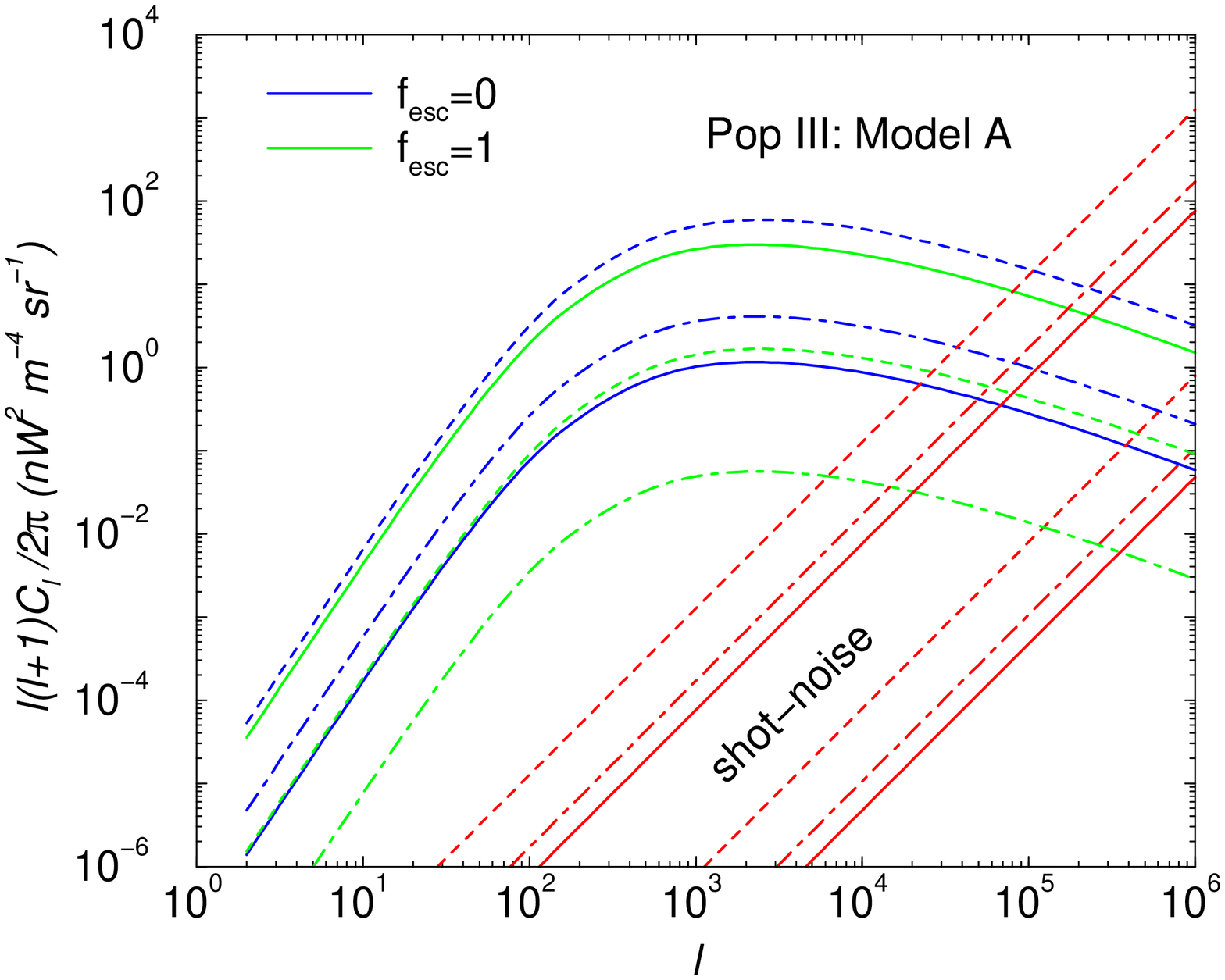,width=2.9in,angle=-90}
\psfig{file=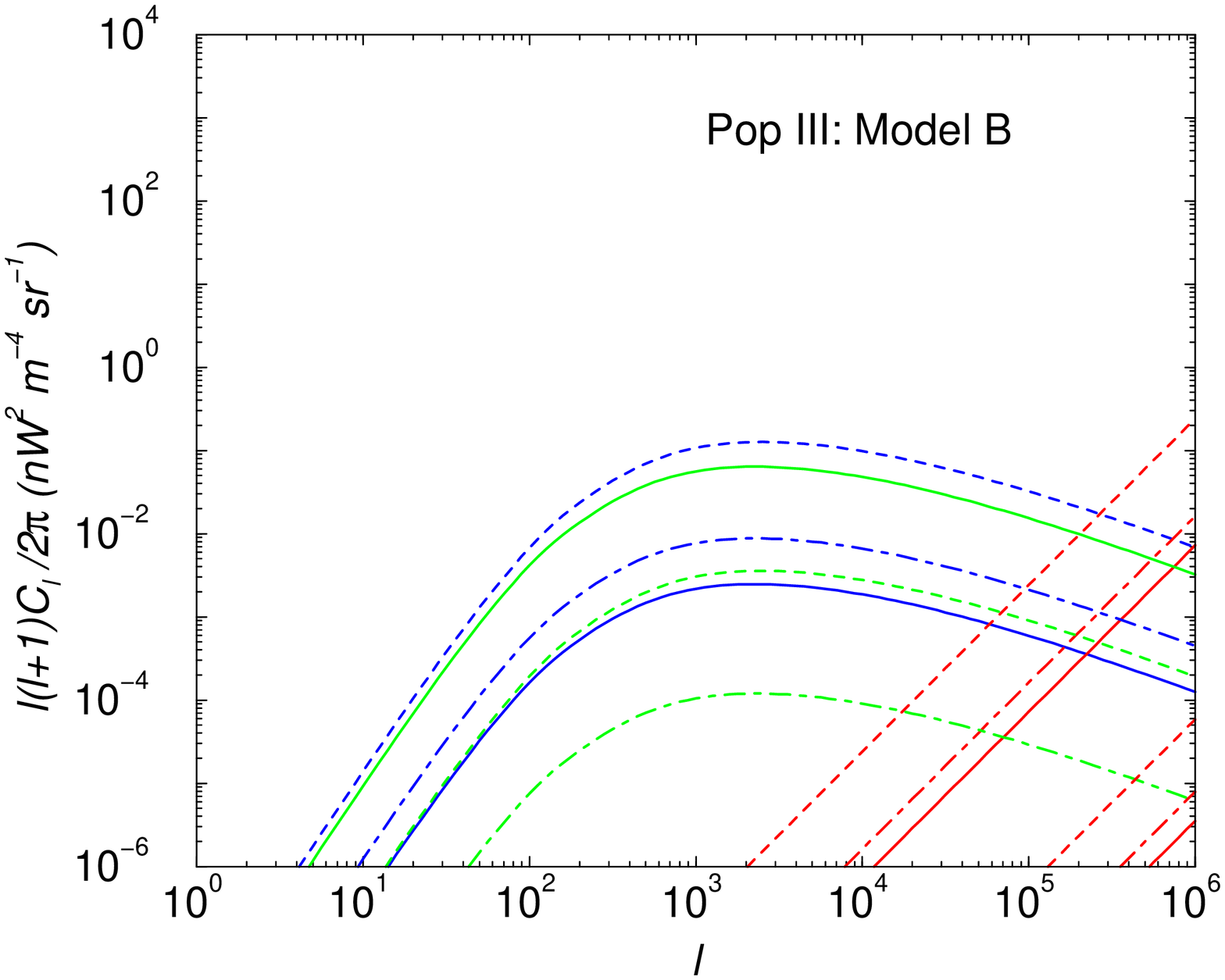,width=2.9in,angle=-90}}
\centerline{\psfig{file=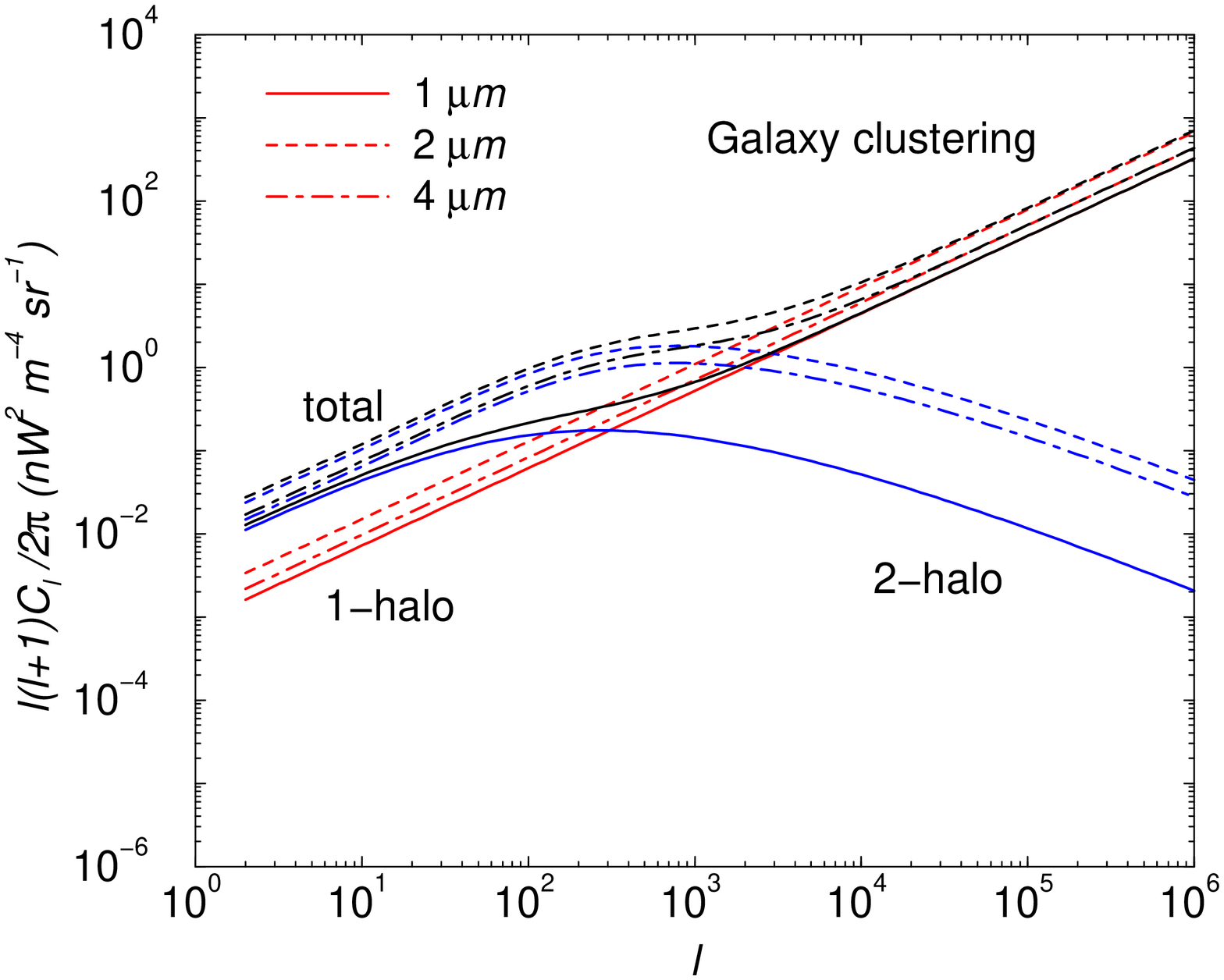,width=2.9in,angle=-90}
\psfig{file=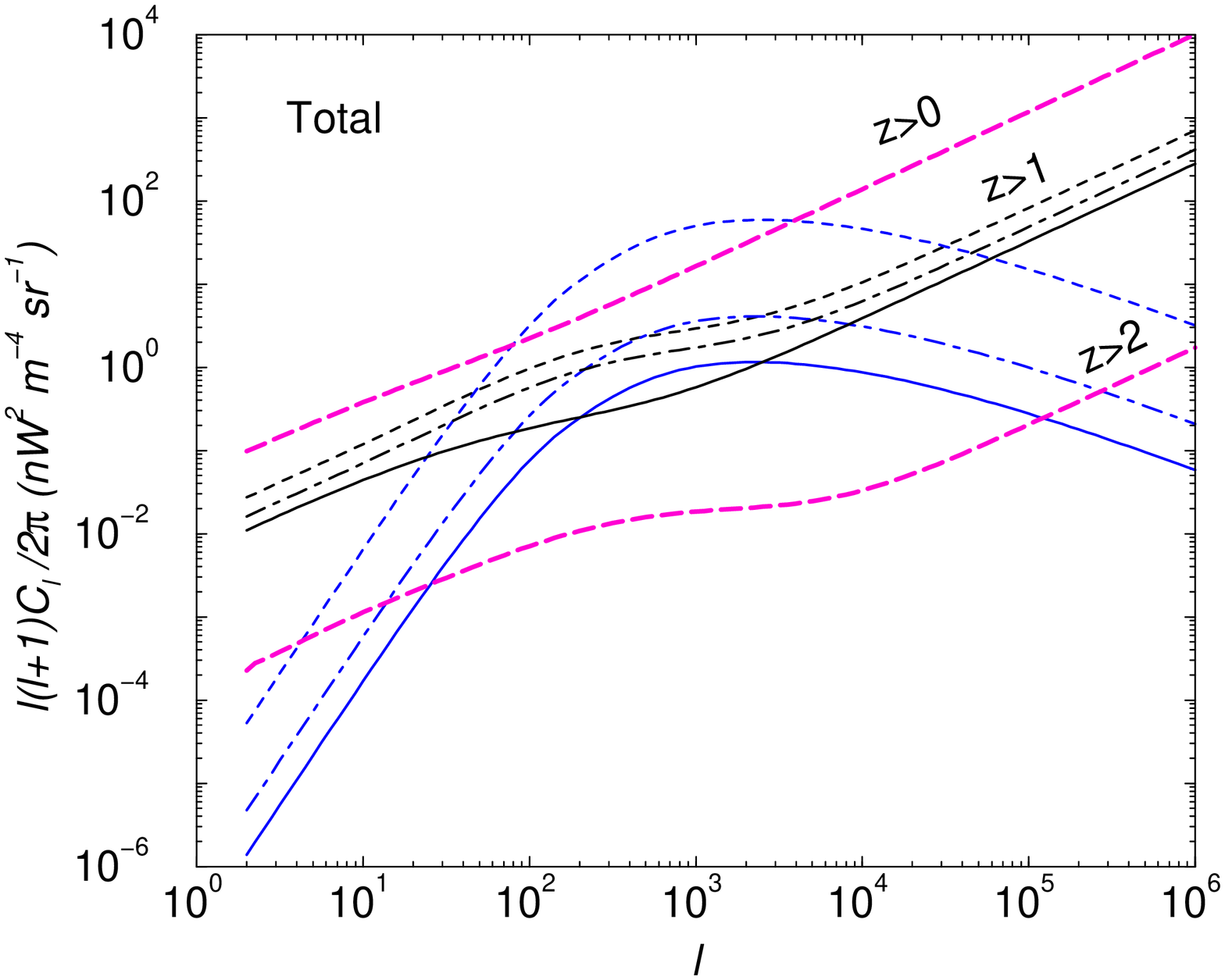,width=2.9in,angle=-90}}
\caption{The angular power spectrum of IRB anisotropies  due to background Pop III stars (top two plots),
and background galaxies (lower plots). The solid, dashed and dot-dashed lines show clustering at 1 $\mu$m, 2 $\mu$m, and
4 $\mu$m, respectively. In the case of Pop III stars, the clustering power spectra are shown for 
two estimates which are at the high (model A; top-left panel) and low (model B; top-right panel) end of the amplitude estimates 
 based on the redshift distribution and biasing factor of dark matter halos, containing Pop III sources, with respect to the linear
density field.  For each of these models, we also plot spectra for the two values of $f_{\rm esc}$, with solid lines showing $f_{\rm esc}=0$
and faint lines showing $f_{\rm esc}=1$, to illustrate wavelength
dependence on the clustering signal. In addition to the clustering part (2-halo part of the power spectrum),
we also estimate the shot-noise contribution related to the finite density of Pop III stars.
Note that this shot-noise contribution related to Pop III stars are highly uncertain due to reasons discussed in the text
and we show the upper and lower ranges of allowed estimates, each in case of models A and B, respectively, in the top two plots.
As shown in the bottom left plot,  the foreground galaxies  have an almost power-law type clustering spectrum; here we only consider
galaxies with redshifts greater than 1 under the assumption that nearby foreground population can be resolved and
removed from the data. Note that we do not show the
galaxy shot-noise spectrum here, but, refer the reader to Fig.~4 for details. We compare galaxy 
and Pop III clustering using model A and $f_{\rm esc}=0$ in the bottom-right plot. As shown, background Pop III stars are expected
to show an excess clustering at tens of arcminute scales due to the strong biasing of dark matter halos that contain these stars at
redshifts of order 15.  In the bottom right plot, for comparison with thick long-dashed lines, we also show the
total (1-halo + 2-halo) galaxy power spectrum at 2 $\mu$m when all galaxies (top curve) and galaxies at redshifts greater than 2 (bottom
curve) are not removed from the imaging data. The extent to which Pop III
clustering can be studied depends on the ability of an experiment to resolve  galaxies at redshifts out at most to 2 
and remove them from the clustering analysis. }
\end{figure*}

We will calculate this power spectrum of the source distribution, related to
both Pop III stars and foreground galaxies,  based on the halo approach (see, Cooray \& Sheth 2002 for a recent review).
In general, the  total power spectrum can be written as a combination of 1- and 2-halo terms with
\begin{eqnarray}
P_{ss}(k) &=& P^{1h}(k) + P^{2h}(k) \nonumber \\
P^{2h} &=&  \left[\int dm\; n(m)\; b(m) \frac{\langle N_{\rm s}(m)\rangle}{{\bar n}_s} u(k|m) \right]^2 P^{\rm lin}(k)\nonumber \\
P^{1h} &=&  \int dm\; n(m)\; \frac{\langle N_{\rm s}(N_{\rm s}-1)(m)\rangle}{{\bar n}_s^2} u(k|m)^p \, ,\nonumber \\
\label{eqn:pk}
\end{eqnarray}
where $u(k|m)$ is the density profile (e.g., NFW profile of Navarro, Frenk \& White 1996) 
in Fourier space normalized appropriately with mass, 
$n(m)$ is the mass function (e.g., PS mass function of Press \& Schechter 1974), and $b(m)$ is the halo bias 
(e.g., Mo \& White 1996; Mo, Jing \& White 1997). The source distribution within halos is encoded by
 $\langle N_{\rm s}(m)\rangle$, the mean source occupation number or the average 
number of individual sources  of interest in each dark matter halo of mass $m$, 
and $\langle N_{\rm s}(N_{\rm s}-1)(m)\rangle$, the second moment of the source distribution. 
In the case of a Poisson-type distribution of sources,
$\langle N_{\rm s}(N_{\rm s}-1)(m)\rangle = \langle N_{\rm s}(m)\rangle^2$. In Eq.~\ref{eqn:pk}, 
$P^{\rm lin}(k)$ is the linear power spectrum of the density field and we use  the
transfer function of Eisenstein \& Hu (1998) to describe the small scale behavior of this power spectrum.
Note that the mean density of sources is given by
\begin{equation}
{\bar n}_s = \int dm n(m) \langle N_{\rm s}(m)\rangle \, .
\end{equation}
Note that this approach to describe clustering of sources that lead to the IRB is similar to the one used in
 Song et al. (2003) to describe clustering properties of far-infrared background sources.

Using the Limber approximation (Limber 1954), the angular power spectrum for a distribution of sources that trace a 
three-dimensional power spectrum $P_{ss}(k)$, when projected on the sky, is given by
\begin{equation}
C_l^{\lambda \lambda'}  = \int dz \frac{dr}{dz} \frac{a^2(z)}{d_A^2} \bar{j}_\lambda(z) \bar{j}_{\lambda'}(z) P_{ss}\left(k=\frac{l}{d_A},z\right) \, ,
\end{equation}
where the comoving angular diameter distance is
\begin{equation}
\da = H_0^{-1} \Omega_K^{-1/2} \sinh (H_0 \Omega_K^{1/2} \rad)\, .
\end{equation}
Note that as $\Omega_K \rightarrow 0$, $\da \rightarrow \rad$. At $l=50$, the Limber approximation is valid to within $\sim$ 5\% of the
exact calculation that involve radial integrations over the spherical Bessel functions (see, Knox et al. 2001 for the formulae that involve the
exact calculation), and rapidly converges to the approximated calculation here as one moves to higher $l$. 
At multipoles below $l \sim 50$, the Limber approximation results in an overestimate of power at the level of 10\%, and we
do not consider this to be important for the present discussion given the large uncertainties, which lead  to several orders of magnitude change in $C_l$,
 associated with the description of Pop III sources and discussed in Section 2.1.

In addition to the clustering signal, at small angular scales, the finite density of sources leads to a shot-noise
type power spectrum in the IRB spatial fluctuations. This shot noise can be estimated through number counts, $dN/dS$, of the contributing
sources, as a function of flux $S$, and can be written as
\begin{equation}
C_l^{\rm shot} = \int_{0}^{S_{\rm cut}} S^2 \frac{dN}{dS} \; dS \, ,
\end{equation}
where $S_{\rm cut}$ is the flux cut off value  related to the removal of resolved sources. 

We will now describe how $\bar{j}_\lambda(z)$ and ingredients related to $P_{ss}(k)$ can be obtained from simple 
analytical methods making use of various approaches in the literature.

\begin{figure*}[!ht]
\psfig{file=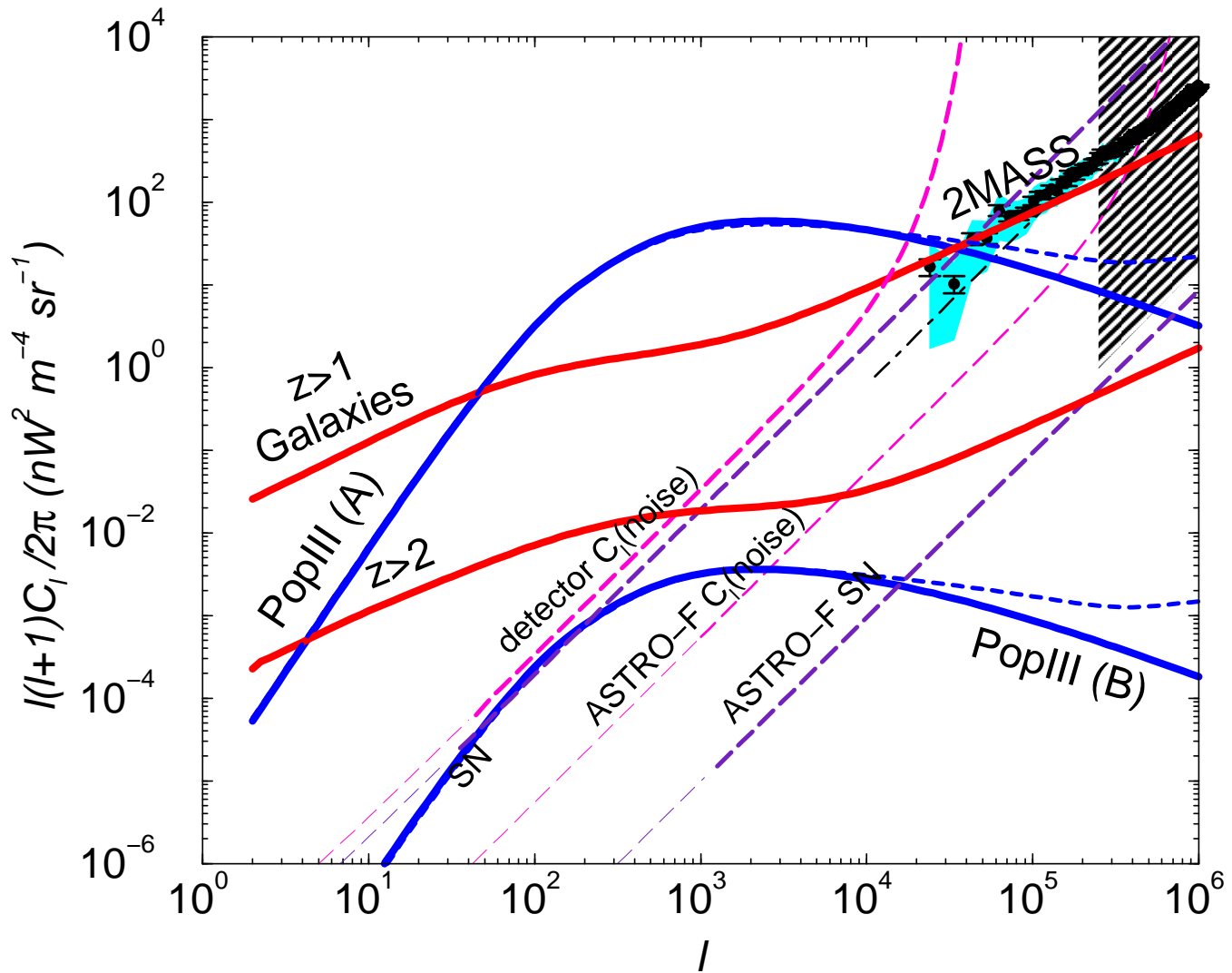,width=7.0in,angle=-90}
\caption{The IRB angular clustering power spectrum at a wavelength of 2 $\mu$m. We show our predictions at this
wavelength based on foreground galaxies (with all galaxies above a redshift of 1) and Pop III stars.
The distinct signature of the Pop III population is at tens of arcminute scales related to linear clustering at redshifts
greater than 10, though its exact amplitude is currently uncertain
by at least five orders of magnitude based on the optimistic (A) and pessimistic (B) models we have considered. 
The dashed line shows an estimate of the non-linear correction to the Pop III angular clustering power spectrum; non-linearities are important
at small angular scales, though, at such scales, we expect the shot-noise do be the dominant source of clustering (see, Fig.~3). 
For reference, we show an average
power spectrum at 2 $\mu$m by Kashlinsky et al. (2003) using  deep 2MASS images. The small shaded region surrounding
2MASS data indicate the field-to-field standard deviation related to the small size of the patch used for these measurements. 
Following K-band galaxy counts (Fig.~2), and the description by Lagache, Dole \& Puget (2003), we have
calculated the shot-noise contribution related to unresolved galaxies below the instrumental sensitivity and show this with a dot-dashed line.
An addition contribution beyond this shot-noise may be needed to explain 2MASS measurements fully, and this could
easily involve Pop III sources as was demonstrated by Magliocchetti, Salvaterra \& Ferrara (2003).
The large hatched region in the upper right corner of the plot show the potential exclusion region for a shot-noise contribution based on a
representative upper limit on the fluctuations in deep K-band galaxy counts (taken to be $K > 24$ at 5-$\sigma$ in a one-arcsecond pixel).
Any reasonable Pop III clustering contribution, with a corresponding Pop III shot-noise power spectrum that passes through this region,
may already be constrained.
For reference, we also plot the instrumental noise power spectra 
in K-band of the planned rocket-borne experiment and ASTRO-F with instrumental characteristics tabulated in Table 1 and 2; the
thick part of these noise curves illustrate the multipoles that will be covered due to restrictions from the field-of-view.
This rocket experiment will be shot-noise limited by the galactic star and foreground galaxy
counts and we show an estimate, after removing point sources at the level of 5 $\sigma$ and above, with another dashed-line labeled ``SN''. 
With 5 $\sigma$ source removal down to a K-band magnitude of 21.3,  the stellar and foreground galaxy shot-noise contribution to the ASTRO-F data
(shown as a dashed line labeled ``ASTRO-F SN'') 
is significantly below the expectation for the rocket-borne experiment and will probe the presence of Pop III clustering down to the
low end of the allowed range in models we have considered.}
\end{figure*}

\subsection{Pop III stars}

To calculate $\bar{j}_\lambda(z)$, we follow the  calculation in Santos, Bromm \& Kamionkowski (2003) and
assume that the mass of stars formed per halo is $m_\star=\eta \Omega_b/\Omega_m m$ when $m \geq m_{\rm min}$ and zero otherwise where
$\eta$ is the star-formation efficiency, which we take here as a free parameter. 
Furthermore, we assume that Pop III stars trace the star formation history at high redshifts and, for calculational purposes,
make use of a model that include  molecular hydrogen cooling at a temperature of  400 K. The star formation rate is calculated as 
\begin{equation}
\psi(z) = \eta \frac{\Omega_b}{\Omega_m} \; \frac{d}{dt} \int_{m_{\rm min}}^\infty dm\; m \frac{dn}{dm} \, ,
\end{equation}
where $M_{\rm min}$ is given in Eq. 4 of Santos, Bromm \& Kamionkowski (2003).
The star-formation rate is plotted in Fig.~7, and we will return to
this later in the context of its estimation from Pop III clustering data in the the IRB spatial fluctuations.

Given the star formation rate, we can write
\begin{equation} 
\bar{j}_\lambda(z) = \frac{1}{4\pi} F_\lambda\;  \langle t_{\rm PopIII} \rangle \; \psi(z)  \, ,
\end{equation}
where $\langle t_{\rm PopIII} \rangle=10^6$ years is the mean life time of a Pop III star and $F_\lambda$ is the stellar flux as a function of wavelength,
including nebular emission, of a Pop III star (Fig.~1). Again, we make use of 
calculations by Santos, Bromm \& Kamionkowski (2003) on the Pop III spectrum
and consider two extreme cases with $f_{\rm esc}=0$ and 1, where $f_{\rm esc}$ is the escape fraction of
ionizing radiation to general intergalactic medium  from the nebula. The spectra are reproduced in Fig.~1
for a star with mass $300$ M$_{\sun}$. While the stellar spectrum can easily be described by a simple black-body,
the nebular emission, related to Lyman-$\alpha$ radiation and free-free emission, involves a detailed calculation and is
model dependent on $f_{\rm esc}$. In the case of $f_{\rm esc}=1$, one finds more emission at shorter wavelengths and we will see later that
this leads to an increase in the clustering amplitude at shortest wavelengths when compared to the case with $f_{\rm esc}=0$.

To calculate the three-dimensional power spectrum of sources, we make use of the same assumptions as above including
the fact that the Pop III occupation number is simply determined by the halo mass such that $\langle N_s(m) \rangle \propto m$ when
$m \geq m_{\rm min}$ and zero otherwise. Since
the number of stars formed is linearly proportional to the halo mass, the Pop III clustering is expected to simply trace
that of dark matter, but with a bias factor determined by the halo masses in which stars are found. 
We also assume a second moment which is given by the square of the mean, as in the case of a Poisson
distribution. At large scales, when $u(k|m) \rightarrow 1$, the simple dependence on the halo mass 
leads to the well known mass averaged bias as
\begin{equation}
\langle b_m \rangle  = \frac{\int_{M_{\rm cut}}^\infty dm \; m\; b(m) n(m)}{\int_{M_{\rm cut}}^\infty dm\; m\;  n(m)} \, ,
\end{equation}
where $b(m)$ is the halo bias with respect to the density field. 

Since our model has a large number of uncertainties,
both in terms of the redshift distribution of Pop III sources, their spectra,  and the biasing factor of halos containing these
stars, we consider a range of models to establish both optimistic  (model A) and pessimistic (model B) estimates of the
angular power spectrum. This is similar to the approach considered in Oh, Cooray \& Kamionkowski (2003) to estimate the
Pop III supernovae contribution to CMB anisotropy fluctuations. 
Our optimistic estimate involves: 
(a) highly biased sources with a cut-off mass for Pop III containing dark matter halos that corresponds to a temperature of
10$^5$ K, (b) a low end for the Pop III redshift distribution at a  redshift of 10, (c) and a star-formation
efficiency of 100\% ($\eta=1$) such that all baryons in these halos convert to Pop III stars.
 Our pessimistic estimate involves: (a) sources which have low bias factors (with a minimum temperature set at 5000 K), (b)
the low end of the redshift distribution of Pop III sources set at 15, and (c) a lower star-formation efficiency with $\eta=0.1$. 
In general, the supernovae related to Pop III stars of model A, following Oh, Cooray \& Kamionkowski (2003), can generate 
the excess CMB anisotropy fluctuations detected by small angular-scale CMB experiments.  In each of these two models, due to the uncertainty
related to the Pop III spectrum, we also vary $f_{\rm esc}$ between 0 and 1 to consider variations allowed for the
amplitude of the angular power spectrum.

For Pop III sources, note that the source power spectrum, in the angular scales of interest, 
is fully described by the linear clustering power spectrum scaled by a bias factor. The one-halo term, 
related to the non-linear part of the power spectrum, is not significant due to the fact that strong biasing
of halos containing Pop III sources leads to a dominant two-halo term.  This is also consistent with the 
rapid reduction in non-linear clustering one expects at redshifts of order 10 and higher
when compared to today. In order to study the extent to which non-linearities may be significant, one can introduce the fully
non-linear dark matter power spectrum based on numerical fits to dark matter simulations (e.g., Peacock \& Dodds 1996).
Since sources are not expected to trace dark matter, especially in the non-linear regime of dark matter clustering,
an alternative approach is to calculate the non-linear correction to source clustering based on the one-halo term related to the halo
approach. This requires knowledge related to the second moment of the Pop III occupation number. Here, as an approximation,
we make use of the Poisson model and assume that the second moment is simply given by the square of the first moment.

In addition to the non-linear clustering, at these small angular scales, what determines the power spectrum 
is the shot-noise associated with the finite density of Pop III sources. To calculate the shot-noise power
spectrum reliably we need a detailed model describing number counts of Pop III sources, including the exact mass function of
Pop III stars within each of the dark matter halos. For the clustering calculation, this information is not required since, in the end,
the clustering amplitude depends only on is the total baryonic mass within each halo converted to stars, and not how this
mass gets divided in to stars. Additionally,
the shot-noise contribution is more uncertain than the clustered component since the shot-noise depends on the rarest Pop III sources
which also happen to be the brightest. Similar to our approach for the clustering signal, we also make both optimistic and
pessimistic estimates of the shot-noise contribution, but for simplicity, we only show the shot-noise related to the $f_{\rm esc}=0$ spectrum
(see, Fig.~3); in cases where the clustering signal is high due to 
higher biasing of rarer halos, we also find a higher shot-noise contribution since such rare halos have a low projected surface 
density. Note that our highest shot-noise estimates, corresponding to the most optimistic estimate on the Pop III clustering 
contribution, can be excluded by fluctuation analyses of already existing deep K-band imaging data. Thus, the recent 2MASS clustering 
measurement at small angular scales, discussed with respect to figure~4, cannot be explained with Pop III shot-noise alone.

\begin{figure}[t]
\psfig{file=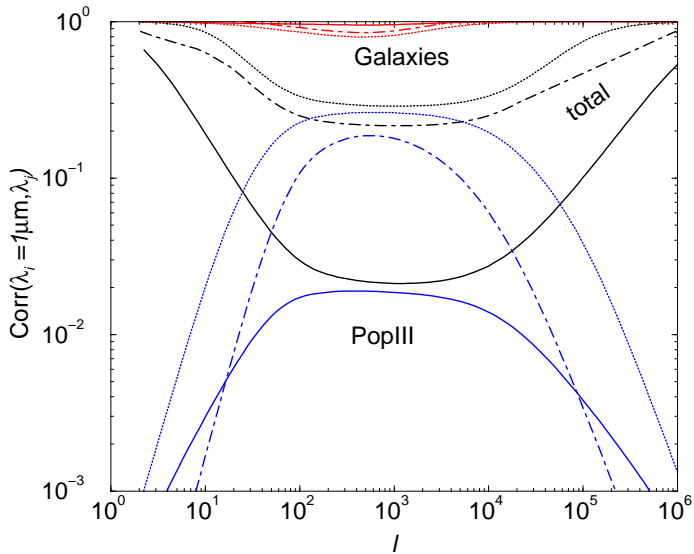,width=3.6in,angle=-90}
\caption{The correlation coefficient of IRB power spectra. Here, we show the cross-correlation
between 1 $\mu$m map and the maps at 2, 4 and 8 $\mu$m, under the assumption that Pop III sources follow our most optimistic description (model A) and
that resolved foreground galaxies are removed out to a redshift of unity. The top lines show the correlation if galaxies were the only contributor
to IRB anisotropies. The Pop III clustering only lead to a correlation at tens of arcminute scales,
which when the two contributions are added, result in an overall reduction of the total correlation coefficient at arcminute scales.}
\end{figure}

\subsection{Foreground galaxies}

To describe the spectrum of an individual galaxy, we make use of the spectrum of M82 (e.g., Silva et al. 1998 where
a model fit is presented to explain the observed spectral energy distribution over a wide range of wavelengths from far-infrared to
UV light). While the use of such a spectrum  to describe all galaxies is likely to be an overestimate, we consider it here since this 
provides a conservative  upper limit on the extent to which foreground galaxies can confuse clustering 
studies of the first star population using IRB data. To describe the redshift distribution of foreground
galaxies, we make use of the normalized star-formation rate between redshifts of 0 and 7, but introduce an exponential
cut off at redshifts above 4 such that the redshift distribution turns over and converges essentially to zero at a value of 7. 
Note that the observational data related to galaxy counts, at redshifts greater than 3, are highly uncertain, though preliminary
studies indicate that the fractional contribution to the infrared background from galaxies at redshifts greater than 3 (and K-band
magnitudes fainter than $\sim$ 23), may not be significant (Totani et al. 2001a). Thus, we expect our modeling to provide only an
estimate of the galaxy contribution to the infrared background.
While a lack of massive halos at redshifts greater than 3 reduces the presence of a large number of IR-bright galaxies considerably,
the presence of unusually red galaxies in certain K-band imaging data (e.g., Totani et al. 2001b; Daddi et al. 2003) 
suggests that the situation may be more complex than suggested by numerical calculations. In addition to the suggested distribution, we 
also considered alternative 
redshift distributions for foreground galaxies, but found consistent results as long as the distribution is
not sharply peaked towards the low redshift end, such as below z of 1; such models are clearly inconsistent with
K-band galaxy counts among other data and we do not consider such models further.

The clustering of foreground galaxies is also calculated following the halo-based approach with a halo occupation number 
based on fits to semi-analytic numerical simulations data by Sheth \& Diaferio (2001):
\begin{eqnarray}
\langle N_{\rm s} \rangle = \left\{\begin{array}{ll}
N_0  & 10^{11}\,M_{\sun} h^{-1}  \leq m\leq M_{B}\\
  N_0\,(m/M_{\rm B})^{\alpha} & m>M_{\rm B}
\end{array}\right.
\label{eqn:galcounts}
\end{eqnarray}
where  we set $M_{\rm B} = 4\times 10^{12}\,M_\odot/h$ and take $\alpha$ and $N_0$ as free parameters.
In our fiducial description, these two parameters take numerical values of 0.8 and 0.7, respectively.
In addition to the mean, we also account for departures from the mean in the 1-halo term with a detailed 
model for the second moment, instead of the simple Poisson description used in above.
Following Scoccimarro et al. (2001), we make use of the binomial distribution, matched to numerical data,
 to obtain a convenient approximation. The second moment is then
\begin{equation}
 \langle N_{\rm s}(N_{\rm s} -1)\rangle^{1/2} =
 \beta(m)\,\langle N_{\rm s} \rangle \, ,
\label{binomial}
\end{equation}
where $\beta(m) = \log \sqrt{m/10^{11} h^{-1} M_{\sun}}$ for
$m < 10^{13} h^{-1} M_{\sun}$ and $\beta(m)=1$ thereafter.
Note that in equation~\ref{eqn:pk},
the simplest approach is to set $p=2$ when calculating $P^{1h}_{dm}(k)$.
In halos which contain only a single galaxy, however, we assume that this galaxy sits at the center and
take $p=2$ when $\left< N_{\rm s} (N_{\rm s} -1)\right>$ is greater than unity and $p=1$ otherwise.
These description for galaxy clustering lead to a power-law like three dimensional spectrum over a wide range of
physical scales and also explains why the clustering of galaxies shows no excess, as in dark matter clustering,
between the transition from linear to non-linear clustering (see, Cooray \& Sheth 2002 for details).

While the use of a low-redshift determined occupation number, motivated by semi-analytical modeling of ``blue'' galaxies,
 to describe clustering properties of IR galaxies may seem problematic, we suggest that this description is adequate for
purposes of the present discussion. The main problem related to our description comes from the fact that, based on observations, one naively expects the
minimum mass at which an IR galaxy at redshifts of $\sim$ 3 to form in halos with a mass at least two orders of magnitude above the minimum mass
we have indicated in Eq. 16. For example, the existing clustering data of IR galaxies at high redshifts (e.g., Daddi et al. 2003)
indicate strong clustering of K-band selected galaxies at a level similar to or greater than the clustering strength of the Lyman
break galaxy population at redshifts $\sim$ 3 and suggests that the minimum mass must be set at a level of $\sim$ 10$^{13}$ M$_{\sun}$.
In the case of Lyman-break galaxies, the halo occupation number is investigated in detail by
Bullock, Wechsler \& Somerville (2002). Contrary to naive expectations, however, these authors find an occupation number with a minimum mass
at the level we have indicated in Eq.~16, though there is a large range of allowed values (see, e.g., Fig.~3 of Bullock, Wechsler \& Somerville 2002).
The difference can be understood based on the fact that the minimum mass, as indicated by the halo occupation number, is not necessarily the
one related to the mass above which halos are found to contain one observed object, but rather the minimum mass in which one galaxy may form.
While analytically, halos with mass below the minimum expected and above the minimum allowed by the halo occupation number form a fraction of a galaxy in
each halo, in reality, this is meant to indicate the fact that only a fraction of such halos form galaxies occasionally. On the other hand, all
halos with mass above the expected minimum mass, which in this case is two orders of magnitude higher than the inferred minimum, are expected to
host one or more galaxies. While the overall dilution of the clustering strength by a sample of small mass halos,
where galaxies are occasionally found, is not significant, these halos are important to describe the clustering properties of the high redshift
population (see, Bullock, Wechsler \& Somerville 2002 for a further discussion).  

We note that our predictions related to IR galaxy clustering at
redshifts around 2 are consistent with Daddi et al. (2003) measurements and find this to be adquate enough for the present paper.
This is mainly due to the fact that we are mainly concerned with clustering properties of brightness galaxies
 at redshifts between 1 and 2, or in the case of K-band, at magnitudes below 20 or so. As we will soon discuss, this is within the range of magnitudes for which
upcoming experiments will be able to remove resolved sources from a clustering analysis.
We are less concerned about the clustering  properties of faint high redshfit galaxies, at $z \sim 3$, for one significant reason.
The suggested measurement involves a clustering analysis of brightness fluctuations and not source counts, as in traditional studies of
galaxy clustering at low redshifts. Thus, the clustering power spectrum from each redshift range, or bin, is weighted by the contribution to the absolute IR
background produced by galaxies in that range or bin. Assuming fluctuations in the brightness trace source counts, 
we can roughly write this as $C_l \sim \bar{I}^2 w_l$ where $w_l$ is the angular power spectrum of source clustering. The fractional contribution to 
IR background from high redshifts is substantially less than that produced by nearby bright galaxies out to a z of 1 (e.g., Cambr\'esy et al. 2001). 
In the K-band, the fractional contribution to the infrared background from galaxies with magnitudes fainter than $\sim$ 20, which are expected to
be at redshifts greater than 2,
is at most at the level of 10\%, or if not less. On the other hand, the fractional contribution from Pop III sources can be at the level of 50\%, or
more, especially under extreme models discussed in the literature (e.g., Salvaterra \& Ferrara 2003). 
In relative terms, using the fractional contribution to the
background, one can safely conclude that Pop III clustering will dominate brightness fluctuations, when compared to that of 
faint galaxies at redshifts of around 3,  by at least a factor of several tens or more since 
$C_l^{\rm  Pop III}/C_l^{\rm  K > 20} \propto [\bar{I}_{\rm Pop III}/\bar{I}_{\rm K > 20}]^2$.  The extent to which this ratio can be known exactly
depends on the relative strength of source clustering at redshifts of around 3 when compared to that at a redshift around 15.
Though some subsamples of faint galaxies, such as the red galaxy (J-K $>$ 1.7) population of Daddi et al. (2003) at redshift of 3,
are strongly clustered, we do not necessarily expect this clustering strength to be significantly higher than Pop III sources such that
$C_l^{\rm  Pop III} < C_l^{\rm  K > 20}$. However, we note that this conclusion could substantially change if the $z \sim 3$ population is to be
a significant contributor to the IR background and its fractional contribution is substantially higher than 10\% or so level implied in the
Cambr\'esy et al. (2001) paper. 

The shot-noise power spectrum related to galaxies, as well as galactic stars, is more certain due to the availability of the number counts
in certain wavelength ranges between 1 and 5 microns. In Fig.~2 (right plot), we plot the average number counts in the K-band
compiled from a variety of sources in the literature following Saracco et al. (2001) and Cambr\'esy et al. (2001; and references therein). 
We will use these number counts to establish the confusion
noise in various experiments, summarized in Fig.~2 left plot, 
at near IR wavelengths to understand the extent to which these foreground sources dominate clustering
studies through their shot-noise power spectrum.  
In both plots of Fig.~2, we also illustrate the magnitude limit to which resolved
sources can be removed at the 5 $\sigma$ instrument noise level in various past and upcoming wide-field imaging data  that are adequate
enough for a clustering analysis of any first-star contribution to the IRB.

\subsection{Angular Power Spectra}

In Fig.~3 (top plot), we show the angular power spectra of galaxies and Pop III sources as a function of wavelength at
1, 2 and 4 $\mu$m. These wavelengths are loosely chosen to represent the clustering behavior of sources in the
IR regime. In the case of galaxies, the angular power spectrum is nearly a power-law over a wide range of angular scales.
The excess clustering at degree scales is related to the linear power spectrum which projects at low redshifts ($z \sim 1$).
When interpreting this galaxy power spectrum, one should consider this as a representative case where foreground galaxies, out to a
redshift of $\sim$ 1, is removed from the data such that nearby low redshift galaxies are absent; In our calculations, we allow
for such a removal by simply cutting off the redshift distribution of galaxies at redshifts below  1.
Note that our approach to describe clustering of galaxies at IR wavelengths is different from previous
approaches. Instead of extending the measured power spectrum at low redshifts, such as from the APM survey,
to higher redshifts, or using numerically calibrated fitting functions for the non-linear power spectrum, such as
from Peacock \& Dodds (1996), which are valid only for dark matter, 
we have made use of the halo approach to calculate the clustering of galaxies directly.

Our approach to describe clustering is also independent of techniques that make use of galaxy number counts at
IR wavelengths. While there is now detailed information related to galaxy counts,
 the exact redshift distribution of IR contributing galaxies still remains somewhat uncertain. As a simple approach, here 
we have made use of a distribution based on the normalized star-formation rate. As long as this distribution is
not dominated by galaxies at the high end of this redshift distribution, 
galaxy clustering follows a power-law.
Though we have not considered here, our halo-based approach can be extended,
 under certain assumptions, to model number counts of IR galaxies
similar to approaches considered in studies such as by Jimenez \& Kashlinsky (1999).

The Pop III stars, since they are present at redshifts greater than 10 and are highly biased with respect to the 
density field, trace the linear regime of clustering at tens of arcminute scales  when projected on the sky today.
As shown in Fig.~3 top plots, the Pop III stars, thus,
show the expected signature of an excess clustering, above the shot-noise power spectrum,
 at  angular scales where the linear power spectrum has a peak associated with the
matter-radiation equality.  The difference between the linear clustering of galaxies at low redshifts and at angular scales
of few degrees or more, and Pop III sources at high redshifts, and at tens of arcminute scales, can be understood based on the
redshift evolution of the linear density field power spectrum alone. As discussed in Cooray et al. (2001), this evolution can
in fact be used to constrain certain cosmological parameters though we do not pursue such possibilities here.

In Fig.~3, we also show Pop III source clustering angular power spectra for flux spectra involving $f_{\rm esc}=0$ and 1.
We consider these two values as a potential bound on the uncertainties related to the Pop III spectrum. 
While the clustering excess at a short wavelength like 1 $\mu$m is small in the case of $f_{\rm esc}=0$,
with $f_{\rm esc}=1$, we find a significant contribution at the level comparable to 2 $\mu$m. This can be understood based on
the Pop III spectra we show in Fig.~1 and results from the fact that when $f_{\rm esc}=1$, the short wavelength cut off is
decreased to a wavelength value lower than for $f_{\rm esc}=0$.  Additionally, due to the decrease in free-free emission
with $f_{\rm esc}=1$, one finds a lower clustering amplitude at high end of the IR wavelengths when compared to the case
with $f_{\rm esc}=0$. While we have considered these two extreme cases
one expects $f_{\rm esc}$ to be order few tens of percent at most, such that more realistic scenario will be more closely 
related to $f_{\rm esc}=0$ than $f_{\rm esc}=1$.

In addition to the spectrum, other uncertainties in physics of the Pop III population
lead to a highly indeterminate overall normalization for the angular power spectrum. These uncertainties, which are more
significant than $f_{\rm esc}$ alone, are related to the Pop III redshift distribution and biasing with respect to the linear density field. 
We bound these uncertainties and consider two estimates at the high and low end; 
it is likely that the true clustering is somewhere between
the two, though, due to large uncertainties, we are unable to determine this to a better accuracy than the range implied
by the two models. While the amplitude of the Pop III angular power spectrum may be uncertain, however,
the angular scale at which the Pop III clustering signature is expected is more reliable as it is simply a reflection of the
linear power spectrum projected at redshifts of order 20; varying the projection from redshifts 10 to 50
or so does not lead to a large change in this angular scale since the angular diameter distance does not change
significantly in this range of redshifts.

In addition to a clustered signal, the finite density lead to a Poisson noise.
These shot-noise power spectra for Pop III sources, as a function of
the wavelength, is also shown in Fig.~3 (top plots). As we discussed earlier, the shot-noise is more uncertain than the
clustering power spectrum since the shot-noise depends strongly on the rare and 
bright events while the clustering amplitude, as well as the cumulative
background, does not.  While the shot-noise 
can easily be confused with the shot-noise power due to faint unresolved galaxies, among other
foregrounds, the clustering excess, at tens of arcminute scales, however, provides a potentially interesting signature of Pop III 
which can be observed.

As shown in Fig.~3 (bottom right plot), the excess Pop III clustering signature lies above the 
angular power spectrum of galaxies when wavelengths are in the range of $\sim$ 1 to 2 $\mu$m and when nearby galaxies
are removed from the data. Allowing for the presence of nearby galaxies leads to a slightly higher
clustering signal for galaxies, which we have shown with a long-dashed line  in the right-bottom plot of Fig.~3 
for a wavelength of 2 $\mu$m. 
The removal of bright nearby galaxies leads to a decrease in the non-linear clustering, but not necessarily the
large angular scale clustering related to the 2-halo part of the power spectrum since the large angular scale
clustering is determined by the linear power spectrum scaled by the bias factor. For galaxy models considered here,
the average bias factor is not strongly sensitive to the removal of nearby sources.
In general, at wavelengths above a few microns, the galaxy
clustering fully dominates IRB anisotropies, though, one can improve the detection of Pop III stars even at these
wavelengths if most resolved galaxies are removed.
For comparison, in Fig.~3, the lower thick long-dashed line show the 
expected clustering in the infrared-background at 2 $\mu$m from galaxies at redshifts greater than 2 suggesting that a removal
down to this redshift range (or magnitudes at the level of 25 in K-band) is essential to extract the clustering
signal associated with first-star population. In general, to perform such a removal and to
detect Pop III clustering signal at tens of arcminute scales,
an experiment with subarcminute-scale angular resolution, but covering several tens of square degrees, is needed.

In Fig.~4,  we highlight the main aspects of Pop III clustering. For reference, we also show the
2MASS clustering measurement in the K-band at small angular scales (Kashlinsky et al. 2003) as a reference.
Note that the Pop III signature is present at tens of arcminute scales with an excess above the level expected from galaxies
(assuming that foreground galaxies, out to a redshifts $\sim$ 1 to 2, can be resolved and removed from the analysis). 
In addition to 2MASS, DIRBE (Kashlinsky, Mather \& Odenwald 1996) and IRTS (Matsumoto 2000)
measured Pop III clustering at large angular scales and detected excess fluctuations (discussed with respect to Fig.~6).
In addition to the linear clustering of Pop III sources, we also show the correction to the power spectrum associated with non-linearities.
While non-linearities is expected to increase the power by an order of magnitude at very small scales ($l \sim 10^5$ to 10$^6$),
a comparison to Fig.~3 reveal that at such small angular scales, the clustering spectrum related to Pop III sources is more
likely to be dominated by their shot-noise. At these same angular scales, the shot-noise related to unresolved foreground galaxies also
become important such that the individual contributions related to galaxies and first stars are not easily separable.

Based on our description of the Pop III population and its clustering aspects,
we find the clustering amplitude resulting from emission by first stars is 
not fully responsible for the clustering signal detected in the 2MASS data at arcsecond angular scales. The 2MASS angular
power spectrum is more likely due to a combination of the non-linear clustering of unresolved galaxies and the shot-noise associated with
such sources.  The number counts of galaxies in the K-band now extend down to a magnitude of 24, and if any shot-noise existed at the
2MASS  level, its fluctuation amplitude from pixel to pixel, due to the Poisson behavior,
 would correspond to a magnitude $\sim$ 22 to 23 in one-arcsecond pixels. The deep counts will easily be sensitive to such variations and
based on such an argument, we suggest that the Pop III source shot-noise or clustering 
contribution must be below the large shaded region on the
top right corner of Fig.~4. This region essentially corresponds to 5 $\sigma$ fluctuations down to a K-band magnitude of 24 in a 
one-arcsecond pixel. On the other hand, using measured galaxy counts alone, 
a significant shot-noise is expected from unresolved sources that are below the instrumental detection limit but
above the confusion limit  (e.g., Lagache, Dole \& Puget 2003 for a description); we show this contribution, together with the
shot-noise from sources below the confusion limit, as a dot-dashed line in Fig.~4. It is likely that a combination of this shot-noise
and the non-linear clustering of the unresolved sources contribute to the 2MASS measurements, though, a substantial contribution from
non-linear clustering of Pop III sources may be expected in certain descriptions of the Pop III population 
(Magliocchetti, Salvaterra \& Ferrara 2003). 

Since various possibilities may exist to explain 2MASS measurements (either in terms of galaxies, Pop III sources or in comibation), 
we suggest that further investigation of small scale clustering behavior of the IRB may not be adequate to separate these
contributions. While one can extend a survey such as 2MASS to large angular scales by tiling subsequent fields, this  can cause
problems when combing adjacent fields to measure large scale clustering beyond that of a single field of view (see discussions in
Odenwald et al. 2003). The alternative approach is to image a wider area (of tens of square degrees) directly at slightly lower resolution 
and perform a clustering analysis, either after removing resolved stars and galaxies directly in such an image or
after accounting for such sources based on higher resolution data of smaller fields, to see if there is any evidence for an excess as indicated by Fig.~4 
related to Pop III sources.  The best combination to search for the Pop III signal appears to be a wide-field camera with
fidelity on degree scales combined with a deep survey, such as ASTRO-F, to minimize foreground confusions.

The detection of excess Pop III source clustering will be limited to the extent that foreground stars and galaxies can
be removed. In surveys where the source removal is not significant, one expects a higher shot-noise
contribution from galactic stars and foreground galaxies, and this shot-noise level could
potentially be above the level of expected clustering of galaxies. To model the expected shot-noise, 
we follow calculations by Lagache, Dole \& Puget (2003), 
 and we show the case for a wide-field rocket-borne experiment (see Table~2 for details) in Fig.~4. The instrumental noise power spectrum
$C_l^{\rm noise}$ related to each experiment is shown with a dashed-line
and is given by
\begin{equation}
C_l^{\rm noise} = 4\pi f_{\rm sky} \frac{\sigma_{\rm pix}^2}{N_{\rm pix}} {\rm e}^{l^2 (\Delta \theta)^2} \, ,
\end{equation}
where $\sigma_{\rm pix}$ is the noise-per-pixel (see, Table~1), $N_{\rm pix}$ is the number of pixels 
and $\Delta \theta$ is the pixel scale or the scale of additional smoothing, if the latter
is employed.  As written, the noise contribution
to the measured power spectra at each wavelength is described through
standard CMB analysis approaches (Knox 1995) given the filtered Gaussian beam and the
noise of the whole focal plane array.
With the noise power spectrum, one can write the error, or standard deviation, for a measurement of
the angular power spectrum at each multiple as
\begin{equation}
\delta C_l = \sqrt{\frac{2}{f_{\rm sky}(2l+1)}}\left[C_l+C_l^{\rm noise}\right]\, ,
\end{equation}
where the term related to $C_l$ accounts for the cosmic variance associated with the finite sky-coverage. When the power
spectrum measurements are binned in the multipole space, the error associated with the binned power spectrum measurement
is reduced further by a factor of $\sqrt{\Delta l}$ where $\Delta l$ is the width of the bin. While we have not shown these
errors in Fig.~4, due to uncertain $C_l$ of Pop III stars, the extent to which upcoming instruments probe the
power spectrum, related to $f_{\rm sky}$ is shown with a combination of thick and thin lines when plotting $C_l^{\rm noise}$.

For instrumental parameter values in Table~1  related to the rocket-based imaging data, one can remove point sources, 
at the 3 $\sigma$ confidence level above the\
 expected instrumental noise and confusion noise from sources below the resolution limit, down to
a flux level of $\sim$ 2 $\times$ 10$^{-7}$ nW m$^{-2}$. This corresponds to a  K-band magnitude of 16.8. Using K-band counts for stars and galaxies,
we expect a shot-noise at the level of $C_l \sim 5 \times 10^{-8}$ nW$^2$ m$^{-4}$ sr$^{-1}$ due to unresolved and unsubtracted sources.
We show this contribution with a dashed line labeled ``SN'' in Fig.~4.   In the case of a wide-field rocket-borne experiment,
one expects a potential detection of the Pop III population, at multipoles of order 10$^3$, if these stars follow
our models at the optimistic end.  

At the low end of our models, the Pop III clustering amplitude is lower and the detection will strongly depend on the extent to which
galaxies, with K-band magnitudes at the level of 20 and fainter, can be resolved and removed from the data.
While reaching such a low magnitude level for resolved sources is beyond the capability of the rocket-based experiment, due to
restrictions in the integration time among others,
such a study is clearly possible with ASTRO-F.  At the 5 $\sigma$ confidence level above instrumental noise,
ASTRO-F allows identification and removal of foreground sources
 down to a magnitude limit of 21.3 in the K-band or 2.2 $\times$ 10$^{-9}$ nW m$^{-2}$ in 1.4 arcsecond-pixels. 
Using the same counts as before, we now find a shot-noise
contribution at the level of $C_l \sim 2.0 \times 10^{-9}$ nW$^2$ m$^{-4}$ sr$^{-1}$; while the extent to which
Pop III star clustering can be detected with ASTRO-F depends on the clustering nature of unresolved galaxies below its detection
limit, we find that ASTRO-F is useful for the study of Pop III stars since it allows a large range in the clustering amplitude to be probed
at arcminute scales due to the limited field-of-view, 10 arcmins by 10 arcmins.
Such an angular power spectrum measurement is  useful to
understand the extent to which clustering of Pop III sources contributes to spatial fluctuations of the IRB given that we predict the
presence of an excess at angular scales close to 10 arcmins and above. In addition to the limited field-of-view, another drawback with
ASTRO-F is the limited wavelength coverage restricted to K-band and longer. To properly establish the clustering of Pop III sources,
one will require a study of the frequency spectrum related to the clustered component and lack of observations below K-band
may complicate such an analysis; Furthermore, if coverage at shorter wavelengths existed, one can use such data for an extraction of
the Pop III contribution, based on its expected spectrum, such that the confusion between Pop III sources and foreground galaxies
are further reduced.

While observations in a single band, e.g., K-band, is useful to understand Pop III clustering, this can be better achieved with
multiwavelength data as different channels can be combined to study clustering
in each band and between bands. 
In Fig.~5, as an illustration, we show the correlation coefficient related cross wavelength power spectra.
This correlation coefficient is defined as
\begin{equation}
C(\lambda_i,\lambda_j) = \frac{C_l^{\lambda_i \lambda_j}}{\sqrt{C_l^{\lambda_i \lambda_i}C_l^{\lambda_j \lambda_j}}} \, ,
\end{equation}
under the assumption that the Pop III sources follow our most optimistic assumptions (model A) and that the foreground
galaxies are removed out to a redshift of 1. As shown, the total correlation coefficient ranges near unity  at large angular scales, decreases 
at tens of arcminute scales, and increases back to almost unity
at small angular scales. If the Pop III contribution were to be non-existent, the correlation coefficient would be more
uniform while the decrease in the correlation at arcminute scales is associated with the Pop III population.
While an individual angular power spectra measured at each wavelength was expected to show excess clustering due to Pop III stars, 
the correlation coefficient from cross-power spectra is expected to show a decrease at the same angular scales.

In addition to separating foreground galaxies and background first-star contribution to spatial clustering in the infrared background,
multiwavelength observations are desirable to understand additional contaminants, among which zodiacal light is important.
While spatial clustering properties of zodiacal light, or dust particles responsible for this emission especially at degree angular-scales,
is unknown, one expects a distinct spectral dependence to its emission given the size-distribution of dust particles. 
With observations at several or more frequencies in the IR band, one can make use of this expected spectral shape of the zodiacal emission
to partly remove its contribution and to preform a clustering study related to the zodiacal emission separately. This is similar to
what has been done in estimating the total IR background in imaging data such as with DIRBE (e.g., Kashlinsky et al. 1996).
With imaging data such as from a rocket, another approach is to make use of the temporal variations related to the zodiacal emission. 
Since the zodiacal light is associated with
micron-size particles near the Earth's orbit, observations spread over the Earth's orbit that span few months or more 
are expected to view a different  dust content towards the same background sky-area at different observational epochs.
This could give additional information to remove the zodiacal light, though, we emphasize  that 
the best option to remove this contaminat is to perform multiwavelength imaging data over the IR band.

As a general consistency check on our models, instead of calculating the angular power spectrum, we also calculated the mean or the
absolute background to be compared with observed results. Note that our Pop III redshift distribution follows that of Santos, Bromm \&
 Kamionkowski (2003) and since we use the same stellar spectrum as they calculated, the Pop III star model produces the absolute
background consistent with Fig.~7  of their paper, to the extent that star-formation efficiency is taken here as a free parameter, and by
varying the minimum redshift at which Pop III sources are found to 7 from 10; in our fiducial description of the Pop III population, we
use a some what higher minimum redshift for Pop III sources and the IR background related to this model results in an overall decrease in the 
Pop III contribution at wavelengths below $\sim 1.5$ $\mu$m. 
Our use of a higher minimum redshift for Pop III sources is motivated by arguments related to
the expected rapid transition from Pop III to Pop II sources, though, we note that based on calculations for Pop II star spectra by 
Tumlinson \& Shull (2000), Pop II stars will continue to make a substantial contribution to the IR background and background fluctuations at
lower wavelengths. In the case of foreground galaxies, our model produces less than $\sim$ 40\% of the background at 2 $\mu$m; this 
is consistent, though potentially higher than, estimates based on galaxy number 
counts (Cambr\'esy et al. 2001). Note that a significant fraction of this contribution comes from
sources at redshifts below $\sim$ 2 and, as expected from studies involving source counts, we expect galaxies out to redshift $\sim$ 2
to be the dominant confusion for a study such as this and the extent to which such galaxies can be removed from the data will determine the
ability to extract Pop III clustering excess in the surface  brightness fluctuations.

In order to consider the extent to which our clustering predictions are consistent with prior results,
we compare our angular power spectra based on  predictions with COBE DIRBE measurements, as a function of wavelength,
 by Kashlinsky, Mather \& Odenwald (1996).
These authors presented a measurement of $C(\theta=0)$, the correlation function at zero lag, or rms fluctuations at the beam scale,
 at several wavelengths and using a top-hat filtered maps. This quantity can be
calculated from our power spectra using the fact that
\begin{equation}
C(0) \equiv \langle \delta I_\lambda^2 \rangle_\sigma = \int \frac{d^2\vecl}{(2\pi)^2} C_l^{\lambda \lambda'} W^2(l\sigma) \, ,
\end{equation}

where $W(x)=2 J_1(x)/x$ is the top-hat window function and $\sigma=0.46^\circ$ is the angular scale of filtering.
We summarize our  results in Fig.~6, where we plot $\sqrt{C(0)}$ for the three cases involving galaxies and Pop III sources (with
$f_{\rm esc}=0$ and 1), and a comparison to COBE DIRBE 
measurements at 1.25, 2.2, 3.5 and 4.9 $\mu$m. 
Note that, in general, our model predictions, even at the optimistic end for Pop III sources,
suggest at least an order of magnitude lower values for $\sqrt{C(0)}$ than measured. The prediction based on galaxies alone
is similar and is consistent with previous estimates (e.g., Kashlinsky  et al. 1996).  
While we have some freedom to fit rest of the data by  varying our model parameters, 
we have not attempted to perform such an analysis here. The Pop III sources, if at the low end of our prediction,
suggest roughly three orders of magnitude smaller contribution to clustering seen in DIRBE data.

For comparison, in Fig.~6, we also show results related to the fluctuation analysis in IRTS data as a function of the wavelength in the IR regime.
These fluctuation measurements are made at the IRTS beam scale of 8 arcmins by 20 arcmins, which probes slightly smaller angular scales
than the DIRBE measurements shown in Fig.~6. There is considerable agreement, however, between DIRBE measurements and IRTS.
It is unclear if these excess fluctuations are due to the shot-noise
associated with unresolved galactic stars, due to the large beams of these experiments, or an additional extra-galactic component.
Note that these observations probed large angular scales, while a substantial contribution from the Pop III population is
expected at tens of arcminute scales and, thus, these data do not necessarily aid in understanding the extent to which first stars are an
important source of spatial fluctuations in IR wavelengths.
In future expriments attempting to study Pop III clustering, sufficient angular resolution to measure and remove bright stars
and galaxies is essential. 

\begin{figure}[t]
\psfig{file=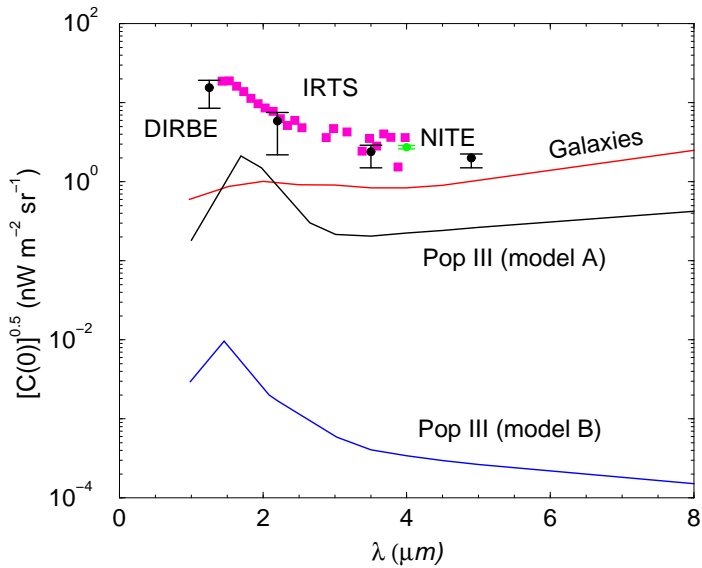,width=3.6in,angle=-90}
\caption{The correlation function at zero-lag and a comparison to DIRBE results from Kashlinsky, Mather \& Odenwald (1996), 
IRTS results by Matsumoto et al. (2000, 2001), and NITE fluctuation measurement by Xu et al. (2002); 
note the difference in angular scales at which these measurements are made: 
DIRBE measurements are at a filtering scale of 0.46$^{\circ}$, IRTS 
measurements  are at the beam scale of  8 arcmins by 20 arcmins and the NITE measurement is at an angular scale  of 17 arcseconds.
Except at 2 $\mu$m, note that our model predictions based on galaxies are roughly an order of magnitude lower than the
measurements. The expected contribution from Pop III stars span at least three orders of magnitude and are highly uncertain; here, we show the
two ranges implies by our data with model A (top line), with $f_{\rm esc}=0$, and model B (bottom line), with $f_{\rm esc}=1$.}
\end{figure}

\begin{table}[hbt]\small
\caption{\label{table:channels}}
\begin{center}
{\sc Instrumental Characteristics: ASTRO-F}\\
\begin{tabular}{ccccc}
\tableskip\hline\hline\tableskip
Band & $\lambda$ & $\Delta \theta$ & $\delta$ F & $\sigma_{\rm pix}$ \\
 & ($\mu$m) &  (arcsec) & $\mu$Jy (5-$\sigma$) & (nW m$^{-2}$ sr$^{-1}$)\\
\tableskip\hline\tableskip
 K & 1.8-2.7 & 1.46 & 1.3 & 14.5 \\
 L & 2.7-3.7 & 1.46 & 1.2 & 15.6 \\
 M & 3.7-5.1 & 1.46 & 1.9 & 17.6 \\ 
\tableskip\hline
\end{tabular}\\[12pt]
\begin{minipage}{3.3in}
NOTES.---%
The sensitivity, $\delta$ F, and noise-per-pixel, $\sigma_{\rm pix}$, is based on 500 seconds of integration per pointing.
The Near Infra-Red Camera (NIRC) of the ASTRO-F mission contains arrays of 512 $\times$ 412 pixels with an effective
area of 10$^2$ arcmin$^2$ per pointing. See, Watari et al. (2000) for more details.
\end{minipage}
\end{center}
\end{table}

\begin{table}[hbt]\small
\caption{\label{table:rocket}}
\begin{center}
{\sc Instrumental Characteristics: Rocket-Borne Experiment}\\
\begin{tabular}{ccccc}
\tableskip\hline\hline\tableskip
Band & $\lambda$ &  $\Delta \lambda/\lambda$ & $\delta$ F& $\sigma_{\rm pix}$ \\
& ($\mu$m) & & $\mu$Jy (5-$\sigma$) & (nW m$^{-2}$ sr$^{-1}$)\\
\tableskip\hline\tableskip
J & 1.25 & 0.24 & 530 & 7.5\\
H & 1.65 & 0.17 & 325  & 6.5 \\
K & 2.2 & 0.16 & 205 & 4.4 \\
L & 3.5 & 0.26 & 134 & 1.8\\
\tableskip\hline
\end{tabular}\\[12pt]
\begin{minipage}{3.3in}
NOTES.---%
The sensitivity, $\delta$ F, and  noise-per-pixel, $\sigma_{\rm pix}$, assumes an useful integration time of 200 seconds.
The pixel size $\Delta \theta$ is 15$''$.'
With pixel array formats of 1024 $\times$ 1024, observations cover a field of view of 4.27 degrees.
\end{minipage}
\end{center}
\end{table}

\section{Reconstructing the Star Formation History}

Since the Pop III redshift distribution is expected to closely follow the star formation history at redshifts between 10 and 30 or so, the anisotropy power spectrum of the IRB can be used to extract information on the 
star formation history, under the assumption of certain aspects related to the source clustering and the spectrum.
For this purpose, we make use of multi-frequency measurements including the cross power spectra between frequencies.
To understand the extent to which star formation history can be studied with IRB fluctuations, we make use of the
Fisher information matrix 
\begin{equation}
{\bf F}_{ij} = -\left< \partial^2 \ln L \over \partial p_i \partial p_j \right>_{\bf x} \, ,
\label{eqn:likelihood}
\end{equation}
whose inverse provides the optimistic covariance matrix
for errors on the associated parameters (e.g., Tegmark et al. 1997).
In Eq.~\ref{eqn:likelihood}, $L$ is the likelihood of observing data set ${\bf x}$, in our case
the angular power spectrum, given parameters $p_1 \ldots p_n$ involved in describing this data.
Respecting the Cram\'er-Rao inequality (Kendall \& Stuart 1969),
no unbiased method can measure the {\it i}th parameter
with standard deviation  less than $({\bf F}_{ii})^{-1/2}$ if all other parameters
are known exactly, and less than $[({\bf F^{-1}})_{ii}]^{1/2}$
 if other parameters are estimated from the  data as well.

\begin{figure}[t]
\psfig{file=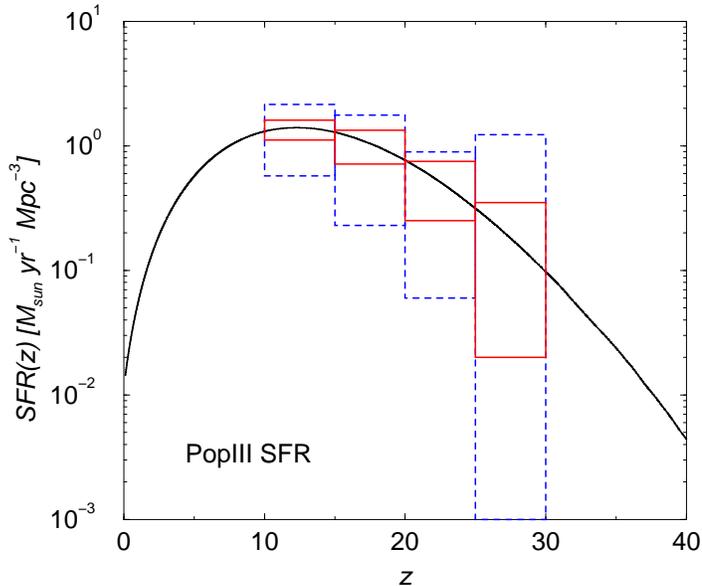,width=3.6in,angle=-90}
\caption{The expected errors on the reconstruction of star formation rate 
based on IRB anisotropy angular power spectra measured with the proposed rocket-borne experiment. 
Here, we assume prior knowledge on the Pop III spectrum and the source bias; in the absence of knowledge on tgese
parameters, one can constrain this combined quantity with the same fractional accuracy as shown in this figure.
In general, the source spectrum may be considered to be known a priori due to the simplistic nature of first stars.
The inner errors show the reconstruction assuming no foreground noise contribution, while the outer errors show
a pessimistic case where we assume an additional shot-noise like power spectrum for the foreground noise,
with a rms at the pixel scale of 10 nW m$^{-2}$ sr$^{-1}$.}
\end{figure}

For the present case involving $C_l$'s of the IRB, we can write
\begin{equation}
{\bf F}_{ij} = \sum_l (l+1/2) f_{\rm sky} {\rm Tr} \left[{\bf C}_l^{-1} \frac{\partial {\bf C}_l}{\partial p_i} {\bf C}_l^{-1} \frac{\partial {\bf C}_l}{\partial p_j} \right]\,,
\label{eqn:Fisher}
\end{equation}
where $f_{\rm sky}=\Omega/4\pi$ is the fraction of sky area covered by the imaging observations.
Here, $\partial {\bf C}_l/\partial p_i$ denotes derivatives of the power spectra matrix, ${\bf C}_l^{\lambda \lambda'}$,  
with respect to the parameter $p_i$ at each multipole and $f_{\rm sky}$ is the fractional sky coverage of the observations. 
This matrix can be written as
\begin{equation}
{\bf C}_l^{\lambda \lambda'} = \sum_i C_l^{\lambda \lambda'(i)} + \delta_{\lambda \lambda'} C_l^{\rm noise} \, ,
\end{equation}
where index $i$ labels all contributions to the power spectrum and $C_l^{\rm noise}$ is the instrumental noise contribution to
the observations. Here, we assume that the noise contribution between maps is uncorrelated.
In terms of individual components of the power spectrum (ie. signal), 
we include contributions from the Pop III stars (which is considered to be the primary component of
astrophysical interest) and the foreground galaxies. 
The clustering properties of zodiacal light is unknown and may be a major source of uncertainty in studies of
early star formation with IRB fluctuations, but due to the unknown properties related to its spatial clustering, we do not
consider its impact in this estimate. More over, the proposed multiwavelength data may allow us to
reduce the confusion related to the zodiacal emission.

In terms of parameters of interest, we study how well ${\bar j}_\lambda(z)$ can be studied with clustering information. We split
this quantity in four redshift bins in the range of 10 to 30 and extract the star-formation rate
in those bins. Note that, in the strictest sense, what one extracts is $\langle b(M) \rangle {\bar j}_\lambda(z)$ which
includes Pop III spectrum information 
and halo bias in addition to the star-formation rate. Here, for simplicity, we assume that both the spectrum and halo bias
 is known though one can constrain a combined quantity instead, if these latter two parameters are assumed to be unknown.
Note the, in general, the spectrum, or at least the shape of the spectrum with an arbitrary normalization,
 can be assumed to be known apriori.

We calculate the Fisher matrix to a maximum $l$ of  5000 such that the clustering of Pop III stars is in
the linear regime (2-halo term) and is not shot-noise dominated (as shown in Fig.~3, we do not expect the Pop III shot-noise
domination to be important until $l \sim$ 10$^4$). In addition to the astrophysical parameters related to the
star-formation rate, we also include cosmological parameters that define the linear matter power spectrum, mainly
the power spectrum normalization, tilt, $\Omega_m h^2$ and  $\Omega_b h^2$. 
We set priors for these parameters as known from current WMAP results  and
its extension related to information from other large scale structure surveys. This information, in general, leads to a power
spectrum at accuracy at the level of few percent and is not a source of major concern for these studies. 

We summarize our results in Fig.~7, where we plot the star formation rate of the Pop III population and reconstructed
errors on this rate based on clustering information for the parameters
as tabulated in Table~1. We assume a total number of 512$^2$ pixels per band
and consider two pointings  with each pointing of 4.3 degrees field of view.
At redshifts or order 10, where most of the Pop III contribution arises, 
the star formation rate is constructed with accuracies of order $\sim$ 0.2 M$_{\rm sun}$ yr$^{-1}$ Mpc$^{-3}$
while at higher redshifts, $\sim$ 30, the clustering information is no longer sensitive to the star formation rate
due to the decrease in fractional distribution of Pop III stars at such high redshifts.

Note that we  have assumed Pop III sources follow the optimistic model 
with a clustering angular power spectrum given by the upper curve shown in Fig.~4.  If clustering were to be lower, then
the reconstructed star formation errors increase. In such a scenario, observations with ASTRO-F can be used to 
investigate the existence of Pop III clustering at arcminute scales to a lower flux level with a better removal of the
foreground galaxies that may confuse the detection of Pop III sources.

\section{Summary}
\label{sec:model}

The recent results related to cosmic microwave background (CMB) anisotropies suggest that the universe was reionized
at a redshift around 17 with an optical depth for electron-scattering of 0.17 $\pm$ 0.04.  Such an early reionization 
could arise through the ionizing radiation emitted by metal-free population III stars at redshifts of 15 and higher.
We discuss the contribution to the infrared background (IRB) surface brightness anisotropies
from such a generation of early star formation. We have shown that the spatial clustering of these stars
at tens of arcminute scales can potentially generate a significant contribution to the angular power spectrum of the IRB.

Note that measurements of IRB spatial clustering already indicate both an overall excess (e.g., Kashlinsky et al. 1996)
and a specific excess at 100 arcmin scales (Matsumoto 2000, 2001) when compared to other angular scales.
The models  based on a contribution from Pop III stars indicate that one should expect a clustering
excess at few tens of arcminutes related to the overall projection of the linear power spectrum at redshifts
between 10 and 30. While the amplitude of the clustering excess is model dependent, the angular scale at which this
clustering is expected is more precise given that it is simply a reflection of the projected linear power
spectrum at redshifts of 10 to 30. The direct detection of the amplitude of clustering excess, as a function of wavelength,  can be
used as a measure of the Pop III starformation rate, under the assumption of a spectrum and clustering bias, or
the combined quantity involving the star-formation rate and bias weighted by the source flux. As an initial attempt in this direction,
we have considered the astrophysical uses of a wide-field camera  optimized for the detection of the first star signature
and have suggested that studies related to Pop III stars can be further improved with near-infrared observations with the upcoming
ASTRO-F mission. The best opportunity to study Pop III stars via IRB spatial fluctuations will come from a combined analysis of
ASTRO-F and  wide-field rocket-borne data in the same area on the sky. The higher resolution data from ASTRO-F can be used to clean 
low resolution, but significantly wider field of view, data from the rocket experiment, such that clustering studies can be extend
even below our pessimistic level of the Pop III clustering considered.

\vspace{0.5cm}

{\it Acknowledgments:} 
This work is supported by the Sherman Fairchild foundation and
DOE DE-FG 03-92-ER40701 (AC), and by a NSF Astronomy and Astrophysics Postdoctoral Fellowship (BK).
AC thanks the Aspen Center for Physics  for hospitality while this work was initiated.
We thank Mike Santos for providing us with Pop III spectra and for useful discussions,
Sasha Kashlinksy for an electronic table of the 2MASS power spectrum and details regarding this measurement,
Aparna Venkatesan for useful comments and suggestions, and an anonymous referee for helpful suggestions.

\end{document}